\documentclass{sigchi}

\toappear{\scriptsize This is an author-prepared print that includes the errata mentioned on the project page at \href{http://visual.cs.ucl.ac.uk/pubs/actionVideo/}{visual.cs.ucl.ac.uk/pubs/actionVideo/}. The original published
version appears in the ACM Digital Library
(\href{http://dx.doi.org/10.1145/3025453.3025880}{dx.doi.org/10.1145/3025453.3025880}).}

\usepackage{balance}  
\usepackage{graphics} 
\usepackage[font=bf]{caption}
\usepackage{subcaption}
\usepackage[T1]{fontenc}
\usepackage{txfonts}
\usepackage[pdftex]{hyperref}
\usepackage{url}
\usepackage{color}
\usepackage{booktabs}
\usepackage{textcomp}
\usepackage{microtype} 
\usepackage{ccicons}  
\usepackage[dvipsnames]{xcolor}
\usepackage{wrapfig}
\usepackage[subtle]{savetrees}

\usepackage{amssymb}
\usepackage{amsmath}
\usepackage[varg]{newtxmath}
\DeclareMathAlphabet{\mathcal}{T1}{pzc}{mb}{it} 

\usepackage{todonotes}

\definecolor{PipelineOrange}{RGB}{247, 148, 30}
\definecolor{PipelineBlue}{RGB}{0, 174, 239}
\definecolor{PipelineRed}{RGB}{190, 30, 45}

\def\plaintitle{Responsive Action-based Video Synthesis}

\def\emptyauthor{}
\def\plainkeywords{Video Editing; Video Textures; Sprites; Cinemagraphs; Interactive Machine Learning.}

\makeatletter
\def\url@leostyle{%
  \@ifundefined{selectfont}{
    \def\UrlFont{\sf}
  }{
    \def\UrlFont{\small\bf\ttfamily}
  }}
\makeatother
\def\pprw{8.5in}
\def\pprh{11in}

\definecolor{linkColor}{RGB}{6,125,233}
\hypersetup{%
  pdftitle={\plaintitle},
  pdfauthor={\emptyauthor},
  pdfkeywords={\plainkeywords},
  bookmarksnumbered,
  pdfstartview={FitH},
  colorlinks,
  citecolor=black,
  filecolor=black,
  linkcolor=black,
  urlcolor=linkColor,
  breaklinks=true,
}

\newcommand{\ch}[1]{{#1}}
\newcommand{\chn}[1]{{#1}}
\newcommand{\chnn}[1]{{#1}}

\usepackage{xspace}
\makeatletter
\DeclareRobustCommand\onedot{\futurelet\@let@token\@onedot}
\def\@onedot{\ifx\@let@token.\else.\null\fi\xspace}
\def\eg{\emph{e.g}\onedot} 
\def\ie{\emph{i.e}\onedot}

\def\wrt{w.r.t\onedot} 
\def\etal{\emph{et al}\onedot}
\makeatother

\begin{document}

\title{\plaintitle}

\numberofauthors{5}
\author{%
  \alignauthor{Corneliu Ilisescu\\
    \affaddr{University College London}\\
    \email{C.Ilisescu@cs.ucl.ac.uk}
    }\\
  \alignauthor{Halil Aytac Kanaci\\
    \affaddr{University College London}\\
    \email{A.Kanaci@cs.ucl.ac.uk}
    }\\
  \alignauthor{Matteo Romagnoli\\
    \affaddr{Testaluna srl}\\
    \email{romagnoli@testaluna.it}
    }\\
  \alignauthor{Neill D.~F.~Campbell\\
    \affaddr{University of Bath}\\
    \email{N.Campbell@bath.ac.uk}
    }\\
  \alignauthor{Gabriel J.~Brostow\\
    \affaddr{University College London}\\
    \email{G.Brostow@cs.ucl.ac.uk}
    }\\
}

\maketitle

\begin{abstract}
We propose technology to enable a new medium of expression, where video elements can be looped, merged, and triggered, interactively. Like audio, video is easy to sample from the real world but hard to segment into clean reusable elements. Reusing a video clip means non-linear editing and compositing with novel footage. The new context dictates how carefully a clip must be prepared, so our end-to-end approach enables previewing and easy iteration.

We convert static-camera videos into loopable sequences, synthesizing them in response to simple end-user requests. This is hard because a) users want essentially semantic-level control over the synthesized video content, and b) automatic loop-finding is brittle and leaves users limited opportunity to work through problems. We propose a human-in-the-loop system where adding effort gives the user progressively more creative control. Artists help us evaluate how our trigger interfaces can be used for authoring of videos and video-performances.
\end{abstract}

\category{H.5.2}{Information Interfaces and Presentation}{User interfaces - Prototyping} 

\keywords{\plainkeywords}

\section{Introduction}
\begin{figure}[t!]
\centering
\includegraphics[width=0.75\linewidth]{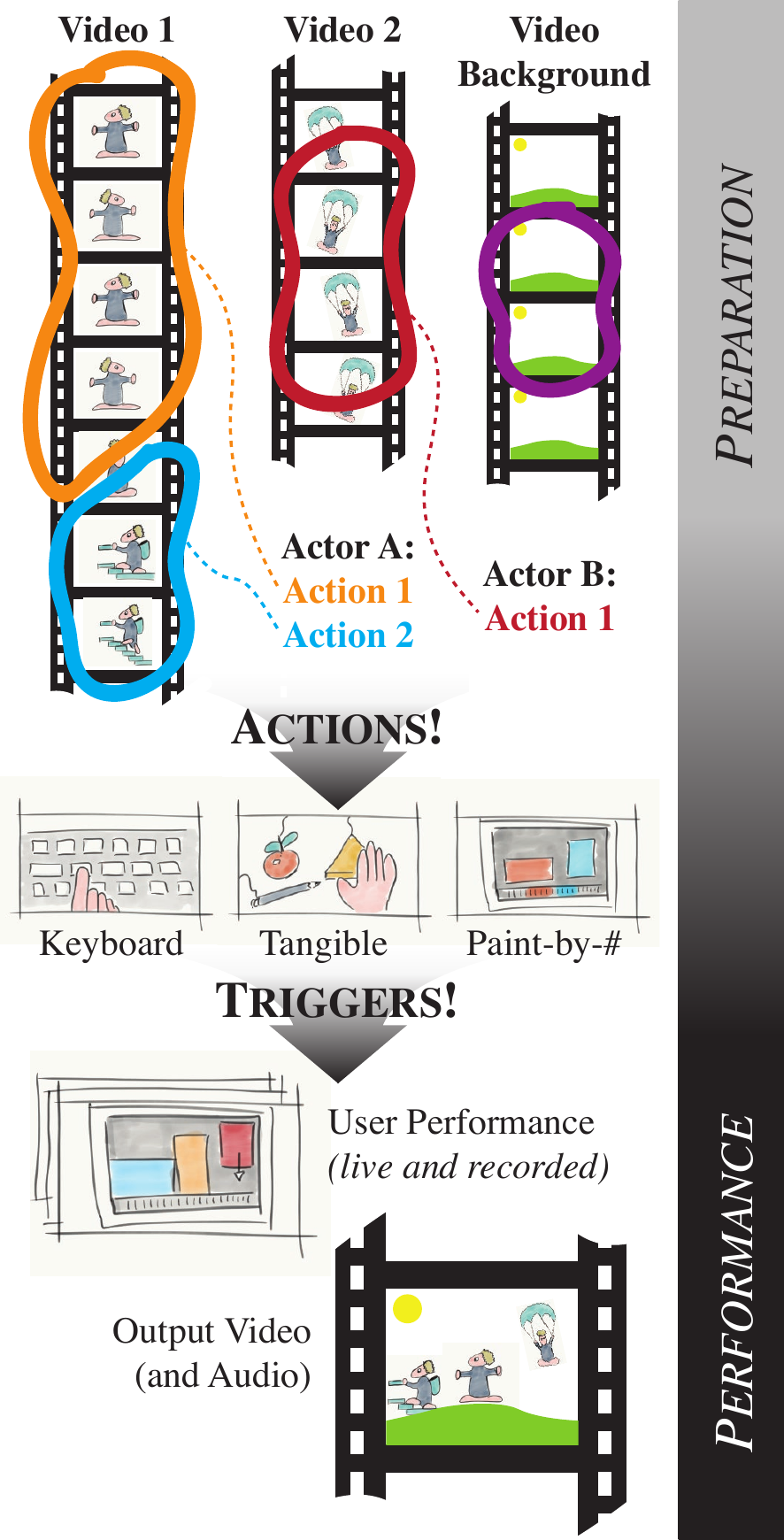}
\caption{Illustration of how \ch{users prepare} raw videos to make an interactive \ch{live} performance. Our end-to-end technology assists with finding and segmenting loopable actions in video inputs (\textcolor{PipelineOrange}{orange}, \textcolor{PipelineBlue}{blue}, \textcolor{PipelineRed}{red}). \ch{Then,} discrete but compatible actions can easily be triggered during a show.}
\label{fig:pipelinefigure}
\end{figure}

In a SIGCHI acceptance speech~\cite{DanOlsenCHITalk}, Dan Olsen outlined the three properties that characterize a great medium of expression: Range, Empowerment, and a Balanced Structure. \chn{Good \textit{range} indicates a wide variety of possible expressions, \textit{empowering} mediums lower the required skills and cost to reach excellent results while a \textit{balanced structure} constrains the user to make new outputs possible.} By these measures, we find Live Looping~\cite{BirthOfLooping}, \chn{where music is recorded and played back in real-time,} to be an inspirational medium for authoring music. Present-day musicians like Reggie Watts\footnote{\url{https://youtu.be/0gKWfvd-chA?t=123s}} and Kimbra\footnote{\url{https://youtu.be/DgmoHtnoi7k?t=27}} can easily accumulate simple sounds,  faithful to the original audio-clips, yet they have the flexibility and precise control to overlay and repeat clips to compose complex music that transcends their solo-musician appearance. We want to make a ``cousin'' of Live Looping for the video domain\footnote{YouTube's MysteryGuitarMan uses labor-intensive methods and has almost 3M followers: \url{https://youtu.be/EQXA7ErL708}}, as illustrated in Fig.~\ref{fig:pipelinefigure}.

\ch{Presently,} technologies for video-authoring have good Range~\cite{DanOlsenCHITalk}, meaning that they are flexible and accurate in depicting many subjects. But they lack a Balanced Structure and Empowerment, which require confining flexibility to ensure even novices succeed, without curtailing what experts can create. \ch{Our goal is to develop a tool that enables a medium of expression characterized by all of these three properties. In particular, we aim to a) ``lower the floor'' so that novices can participate, b) ``raise the ceiling'' so that a single artist can compose expressive pieces and performances while c) catering for the widest range of inputs possible.}

\chn{We achieve this by adapting looping concepts to video in a prepare- and perform-structure (see Fig.~\ref{fig:pipelinefigure}).}
\ch{
In the \textit{preparation} stage, we treat} all moving elements as video \emph{sprites}, \ie a bendy tube of pixels in a stack of sequenced images, like Lu~\etal~\cite{Lu2012timelineediting}. \ch{We call these \textit{actors}.} If a video features only one actor, this is simply a whole-frame sprite.
\ch{Each tube is then manipulated in time, while maintaining the original spatial properties, to create the output. This is done by splitting} a sprite's frames into clusters of \emph{actions}, \ch{and allowing artists to choose which subset of frames to show during the second part of our approach: the live \textit{performance}. A user requests actions through a wide range of trigger interfaces, such as tangible widgets, keyboard, or paint-by-numbers, while our system ensures smooth loops within clusters and transitions between them. Artists can also edit synthesis constraints to ensure sensible actor behavior.}

\ch{
\chn{In this paper, we present the following contributions:}
\begin{enumerate}
	\item a \textit{constrained optimization algorithm} wrapped inside an expressive new interface, that allows users, for the first time, to control actors in video by high level interactions;
	\item an \textit{end-to-end system} to create, iteratively repair, and control video sprites: artists can quickly improve clips that are hard to segment or loop, and check their quality live without jumping between disjoint tool-chains;
	\item a \textit{responsive new medium of expression}, that enables artists to merge video assets
they made and rehearsed earlier in a live performance.
\end{enumerate}}

\ch{We assume the input videos to our system adhere to the following criteria: a) the camera is stationary, b) there are no large differences in lighting over the sequence, c) the background is mostly stationary, d) the filmed \textit{actors} are mostly well separated from each other and e) the \textit{actions} they perform are visually distinct. We show how, despite these assumptions, our system can cater for widely different videos featuring various types of \textit{actors} and \textit{actions} and produce rich and diverse results.} We \ch{also present interviews with} several video-artists and \ch{evaluate the} triggering interfaces to understand the pros and cons of this new medium of expression and adjust the underlying technology.

\section{Related work}
\begin{figure*}[t!]
\centering
	\begin{subfigure}{0.205\linewidth}
    	\includegraphics[height=1.5in]{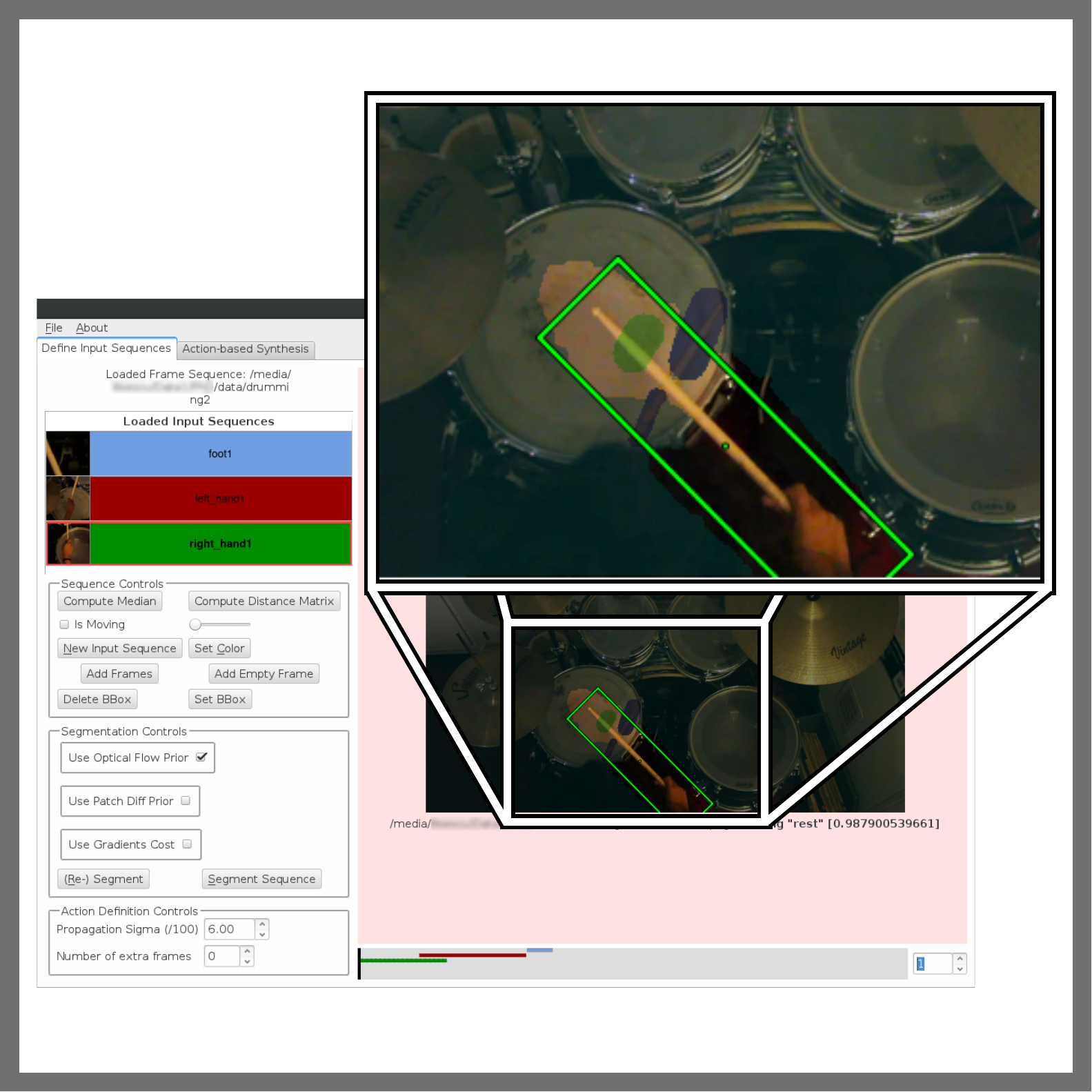}
    	\caption{Track and Segment}
    	\label{fig:overviewtrackseg}
	\end{subfigure}
	\hspace*{\fill} 
	\begin{subfigure}{0.205\linewidth}
    	\includegraphics[height=1.5in]{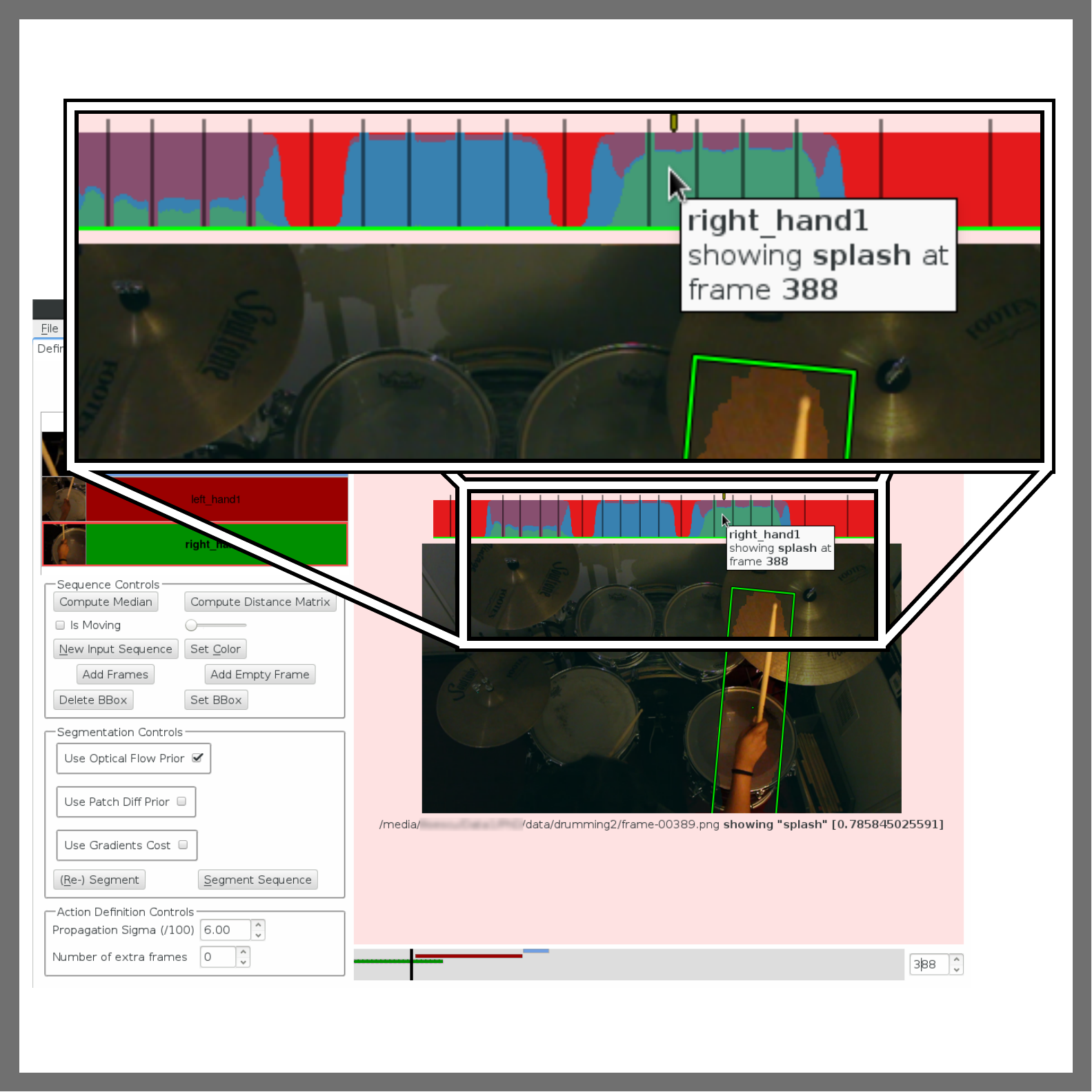}
    	\caption{Action Definition}
    	\label{fig:overviewactions}
	\end{subfigure}
	\hspace*{\fill} 
	\begin{subfigure}{0.375\linewidth}
    	\includegraphics[height=1.5in]{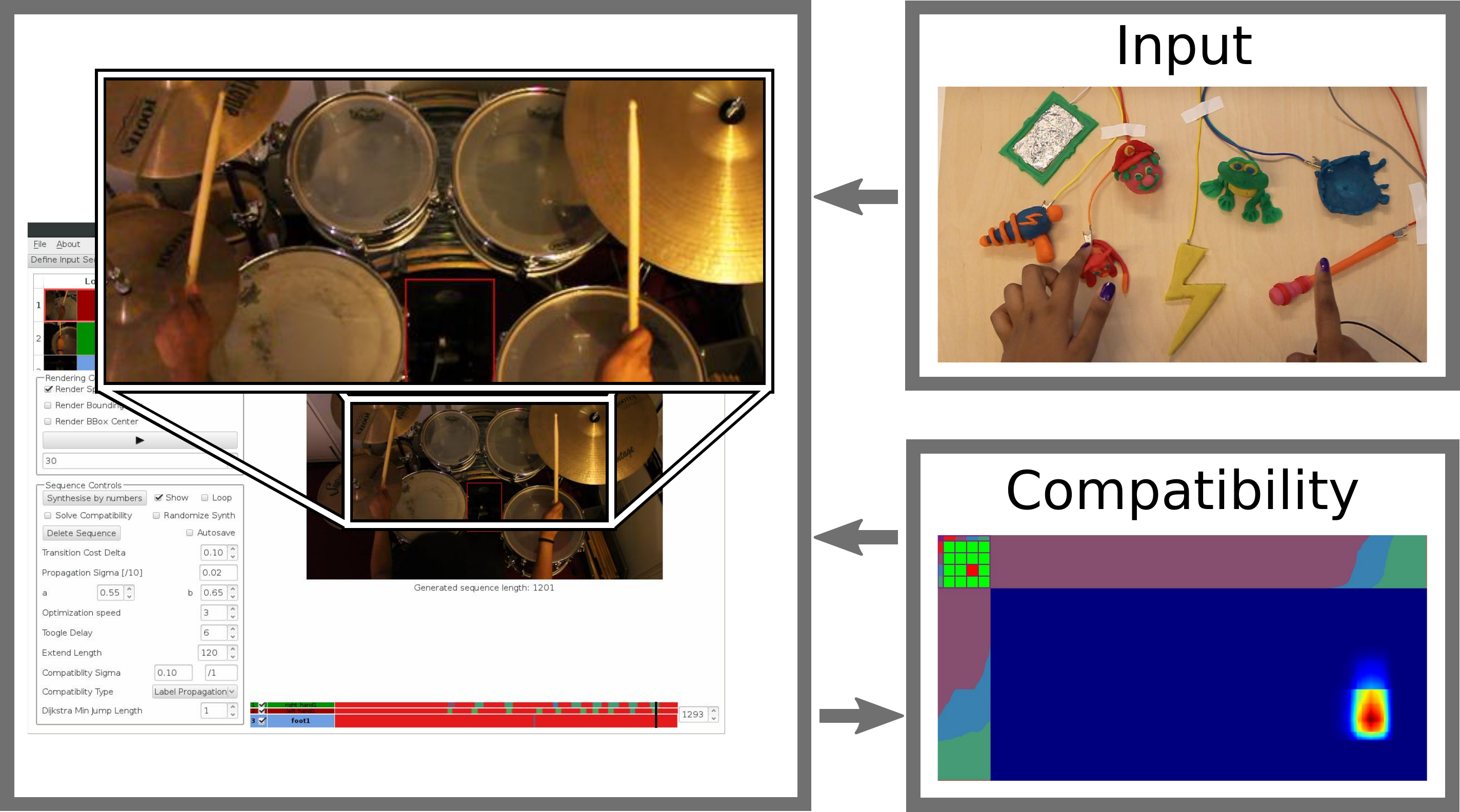}
    	\caption{Responsive Synthesis}
    	\label{fig:overviewsynthesis}
	\end{subfigure}
	\hspace*{\fill} 
	\begin{subfigure}{0.185\linewidth}
    	\includegraphics[height=1.5in]{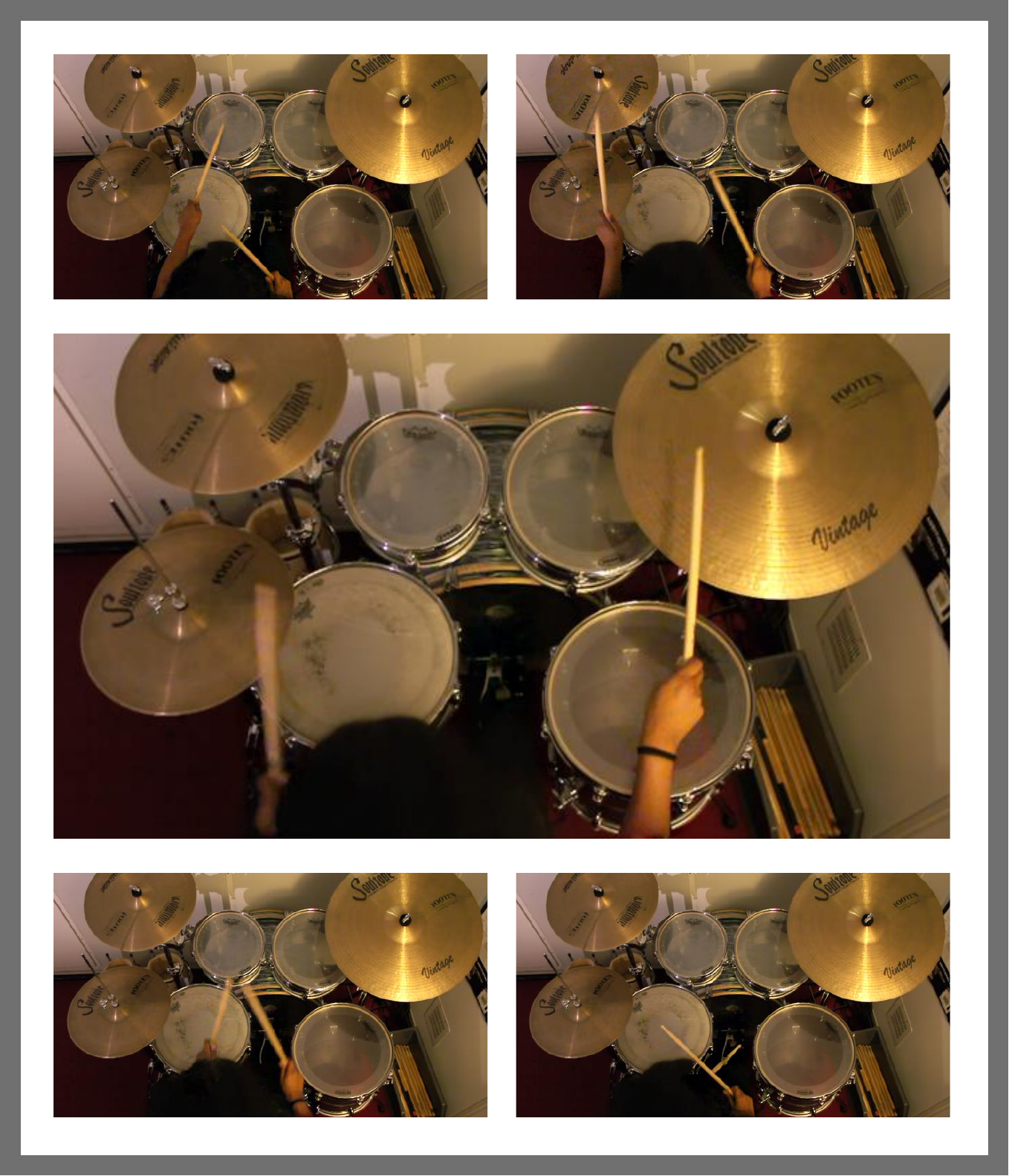}
    	\caption{Rendered Output}
    	\label{fig:overviewrendered}
	\end{subfigure}
\caption{Overview of our interactive video synthesis pipeline: (\subref{fig:overviewtrackseg}) The first, \textit{optional}, step is to track and segment the actors we wish to control, such as the two sticks and foot of the drummer in this example. (\subref{fig:overviewactions}) The user defines a set of actions for each actor by tagging example frames. Here, actions are hitting a specific drum or cymbal and resting. (\subref{fig:overviewsynthesis}) A new video is synthesized given input commands mapped to actions and, optionally, frame compatibility information. The compatibility knowledge is learned over time, as the user tags pairs of frames, and the output is changed accordingly. (\subref{fig:overviewrendered}) The synthesized sequence is composited and rendered seamlessly (using Poisson Blending~\protect\cite{perez2003poisson} and our custom compositing algorithm).}
\label{fig:systemoverview}
\end{figure*}

In this work, we are inspired by methods for direct video manipulation~\cite{goldman2008video} and navigation~\cite{dragicevic2008video, karrer2008dragon}. Eventually, we aim to make interactive synthesis of new videos as fun as \ch{the} manipulation and non-sequential playback of existing videos \ch{is in the above methods. We simultaneously tackle the problems of looping arbitrary videos and influencing synthesis of novel content through user annotation. We now highlight works related to those problems and interesting interfaces for video editing.}

\paragraph{Video looping and animation}
With their pioneering work on \textit{Video Textures}, Sch\"odl~\etal~\cite{schodl2000vt} tackled the problem of indefinitely playing back a finite length video without visible transitions. It worked by finding interchangeable pairs of frames, for single subject videos depicting repetitive or stochastic motion, but struggled with videos containing multiple independent subjects or complex motions.

Kwatra~\etal~\cite{kwatra03gctext} treat video looping as a texture synthesis problem, solved using an energy minimization approach. Extensions for panoramic videos and stereo panoramic videos were proposed \ch{in~\cite{agarwala2005pvt} and~\cite{couture2011} respectively.} Liao~\etal~\cite{liao2013, liao2015seemlessloops} \ch{model motion on a per-pixel basis and segment the input video into spatio-temporal regions of similar motion automatically.} Finally, Sevilla-Lara~\etal~\cite{sevillalara2015smoothloops} \ch{focus} on videos exhibiting camera motion which significantly increases the looping complexity and thus limit themselves to single dominant subjects.

In contrast to the above, our technique \ch{enables} more meaningful synthesis by introducing additional knowledge rather than simply focusing on loop finding and disguising transitions.

\paragraph{Video-based animation}
There are many examples of methods to create novel animations from filmed footage found in the literature. The original \textit{Video Textures} paper by Sch\"{o}dl~\etal~\cite{schodl2000vt} and follow-up work~\cite{schodl2002sprites}, allows segmented sprites of animals, annotated with velocity vectors, to be  controlled interactively using the mouse pointer. In contrast, Bhat~\etal~\cite{Bhat2004flow} create animations of stochastic elements, such as smoke and water, by leveraging user-defined flow patterns to loop and re-position them. Flagg~\etal~\cite{Flagg2009} introduce a specialized technique for human video textures that can create animations of human motions from a database.

The systems described above do not have any knowledge about what is taking place in the input video. Therefore, synthesized video elements look plausible, but random. Some indirect user control is possible by making some assumptions and devising custom energy functions (such as looping through more or less heterogeneous frames~\cite{liao2013}). In contrast, we generalize by giving users the tools to interactively define their subjects and how to control them ``by example''.

\paragraph{Video editing}
\ch{A cinemagraph is a traditionally hand-made medium of expression that combines still and moving imagery. In recent years,} much effort \cite{bai2012sdv, joshi2012cliplets, liao2013, tompkin2011autocine} has gone into \ch{automating this time-consuming} process. Users can decide what areas to animate in~\cite{tompkin2011autocine}, combine small looped clips called \textit{cliplets} in~\cite{joshi2012cliplets} or scribble over patches to automatically animate or de-animate them~\cite{bai2012sdv, liao2013}. Similar to these works, we reduce the degrees of freedom of captured footage by dividing it into disjoint patches. Unlike them, however, we do not restrict users to a binary decision of ``(not) animate'' and allow them to easily decide \textit{how} to animate through our object-action mapping.

Related to our method, video re-timing allows one to re-order filmed events for new and interesting effects. Shah~\etal~\cite{pritch2008synopsis} focus on condensing large amounts of video into short animations but often result in disturbing artifacts such as ghosting. \ch{Users are given} tools to cut and re-arrange trajectories in a spatio-temporal 3D volume in~\cite{Lu2012timelineediting, shah2013videomanip}. In \textit{DuctTake}~\cite{ruegg2013ducttake}, events filmed in the same scene at different times are composed together using a graph-cut energy optimization, while Liao~\etal~\cite{liao2015audeosynth} build on this by allowing music to drive event re-ordering. Finally, Rav-Acha~\etal~\cite{rav2005evolving} bend time for image patches by projecting their pixels onto evolving time fronts.

In contrast to the above techniques, which do not natively support looping, our method can re-arrange frames arbitrarily. Additionally, unlike in our approach, users must synchronize events manually to avoid incompatibilities, \eg colliding cars~\cite{shah2013videomanip}; outputs are limited to ones where filmed interactions are preserved~\cite{Lu2012timelineediting}, or they are not supported altogether~\cite{rav2005evolving}.

\section{System Overview}
\label{sec:system}
We design our end-to-end interface to allow content-creators to quickly prototype their ideas. The more effort they are willing to invest, the higher the quality and complexity their results can \ch{achieve.} Through discussions with six different technical artists, interactivity (as opposed to automation) and responsiveness were identified as stand-out characteristics of this medium of expression; we emphasize these aspects in our prototype system.

Broadly, videos are prepared before being used in one or more performances. With this in mind, we start by providing the necessary tools to define elements of interest which we call \textit{actors}. These can be full-frame video sequences, such as our \textsc{Toy} and \textsc{Candle} datasets (see Fig.~\ref{fig:datasets} and Tab.~\ref{tab:datasetsinfo}), or localized objects, such as the cars in \textsc{Havana} or hands in \textsc{Drumming}. 

For the objects, we provide semi-automatic tracking and segmentation capabilities (Fig.~\ref{fig:overviewtrackseg}). We enable the user to correct any mistakes in the bounding box tracks interactively. Similarly, for separating the tracked object from the background, our tool provides previews of generated action video clips, together or in isolation. Users can then correct and influence the quality of the final segmentation by scribbling over the resulting masks.

The next step is the most critical and represents the core of our new medium of expression. Using our simple UI (Fig.~\ref{fig:overviewactions}), users associate a set of \textit{actions} to each actor, specifying the moment in the video timeline. For instance, each musical note in \textsc{Toy}, or drum hit in \textsc{Drumming}, represents semantically and visually distinct actions. Users define these by tagging a few example frames while the remaining ones are labeled automatically, based on visual similarity, using a machine learning approach. This reduces the required user input and provides almost instant feedback, allowing users to validate the automatic action association and, if necessary, refine it by tagging more examples.

A new video performance synthesizes a number of output layers, each of which corresponds to an actor. Without further guidance, our algorithm can seamlessly loop through the actor input frames by finding visually smooth transitions (similar to~\cite{schodl2000vt}). Users can, however, guide the live video performance by pressing keys mapped to actors' actions (Fig.~\ref{fig:overviewsynthesis}), requesting what to see and when. As we show later, this simple but powerful interaction mechanism enables more creative input methods such as \textit{MakeyMakey}~\cite{makeymakey}, \textit{synthesis-by-numbers}~\cite{hertzmann2001image} or custom game logic. 

Our novel and fast synthesis algorithm balances the importance of meeting users' requests with maintaining the visual quality of loop transitions, to create a new video interactively. Users can further refine the output by tagging incompatible frames or actions, so that actors interact only in desirable ways; for example, diggers should only load parked trucks (see \textsc{Digger} in Fig.~\ref{fig:semanticcompatibility}). Our synthesis algorithm uses this information to improve the resulting output, completing the human-machine feedback loop that makes results possible, in response to high level triggers.

Finally, we can perform an optional post-processing step (Fig.~\ref{fig:overviewrendered}) to improve the quality of the output sequence recorded during the interactive phase described above, producing the final results shown in the supplemental video. We use seamless blending to remove artifacts due to illumination changes and then merge the actor patches together with the background ensuring that the overlapping regions are handled correctly.

The following sections provide the technical and implementation details required to reproduce our system; these are followed by the results and evaluation.

\section{\ch{Actor Preparation}}
\label{sec:preprocessing}
\ch{We now describe the steps and tools used to \textit{prepare} a raw video for use during a live performance. The result of this stage is a set of \textit{actor sequences}: video sprites} associated to actions that can be interactively triggered during synthesis. \ch{Optionally, actors can be tracked and segmented to improve looping and increase output variability.}

\paragraph{Tracking and segmentation}
Critical to looping algorithms is the ability to find similar frames or patches, at different points in the timeline, that can be used interchangeably to ``jump'' between different parts of the video. This is impossible for complex videos, such as ones with multiple, independently moving objects (see \textsc{Havana}). Methods such as~\cite{liao2013} partially address this problem by adapting their patch shape to best suit looping, but are prone to cutting objects, introducing visible seams. We choose to let users decide interactively which elements they may want at showtime.

\begin{figure*}[t!]
\centering
\includegraphics[width=0.82\linewidth]{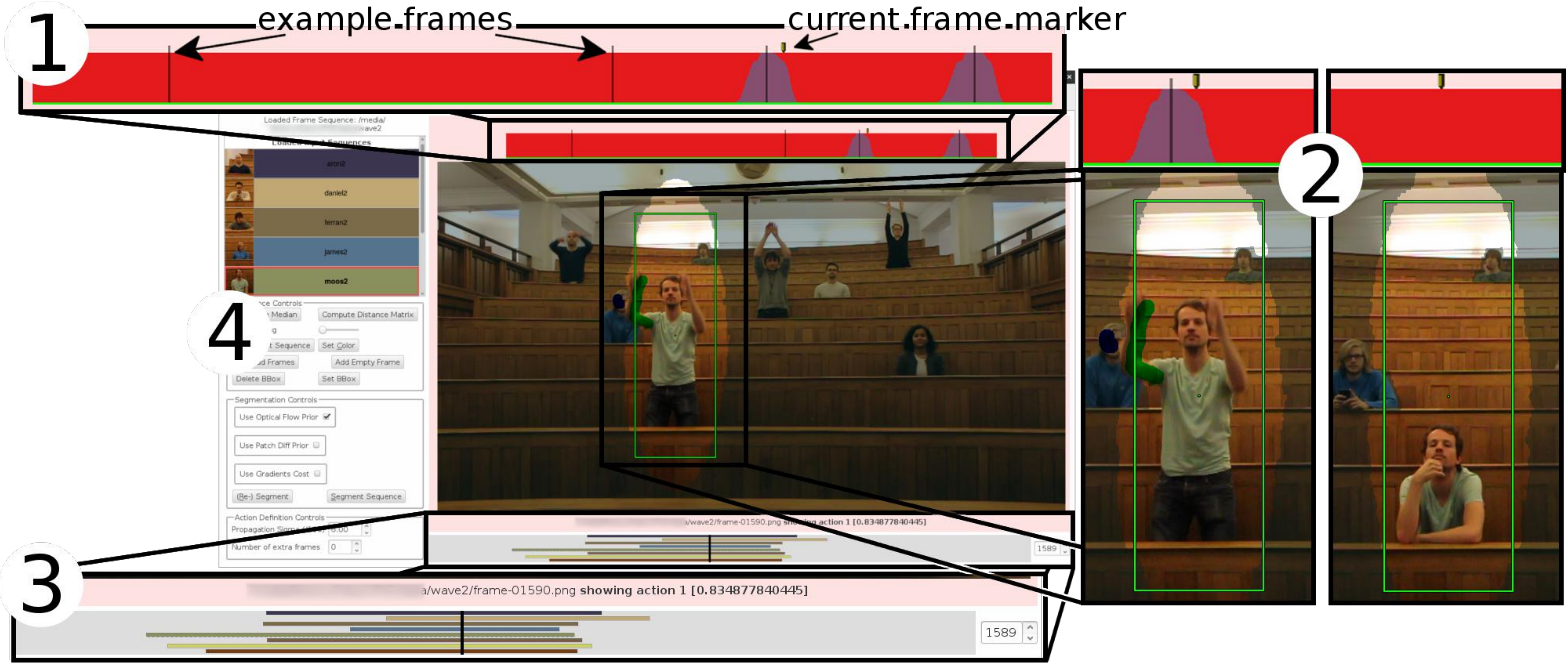}
\caption{Our actor definition user interface: (1) \textit{actions} associated to each frame of a tracked object (\eg the closest person is shown ``sitting'' in each \textcolor{red}{red} frame and ``standing'' in each \textcolor{Fuchsia}{purple} frame, with in-between frames shown as a combination of the two colors); (2) two example frames with associated actions (above each and denoted by marker), bounding box and segmentation with corrective strokes (\textcolor{blue}{blue} = BG, \textcolor{green}{green} = FG); (3) \textit{input video timeline}: the black vertical line indicates the current frame and the colored horizontal lines indicate frames where each actor (in this case people) has been tracked; (4) list of tracked actors (identified by their unique color also used in (3)).}
\label{fig:preprocessingui}
\end{figure*}

First, users track bounding boxes around objects. In our system, we chose to use the CMT tracker~\cite{Nebehay2015CVPR} because a) it is easy and quick to correct in an interactive setting (see our UI in Fig.~\ref{fig:preprocessingui}) and b) it estimates both scale and orientation along with the position of the bounding box.

\begin{figure}[t]
\centering
	\begin{subfigure}{0.205\linewidth}
    	\includegraphics[height=1.1in]{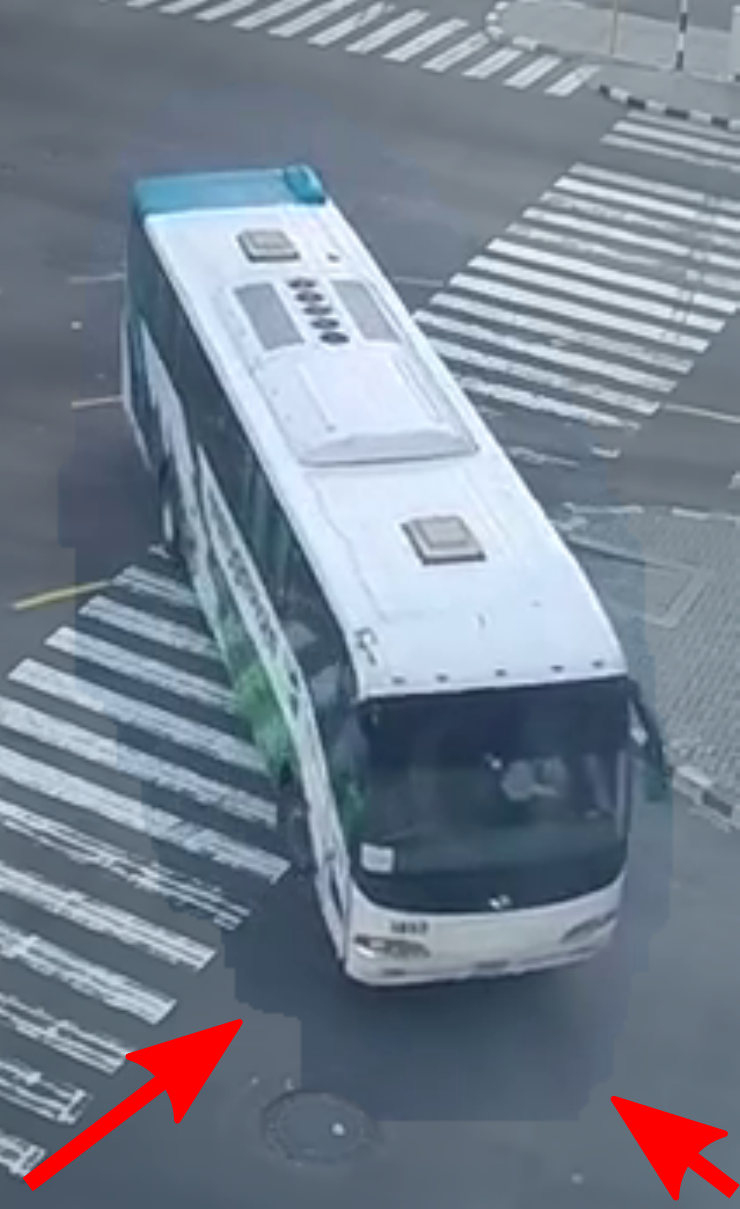}
    	\caption{On BG}
    	\label{fig:patchonbg}
	\end{subfigure}
    \begin{subfigure}{0.275\linewidth}
    	\includegraphics[height=1.1in]{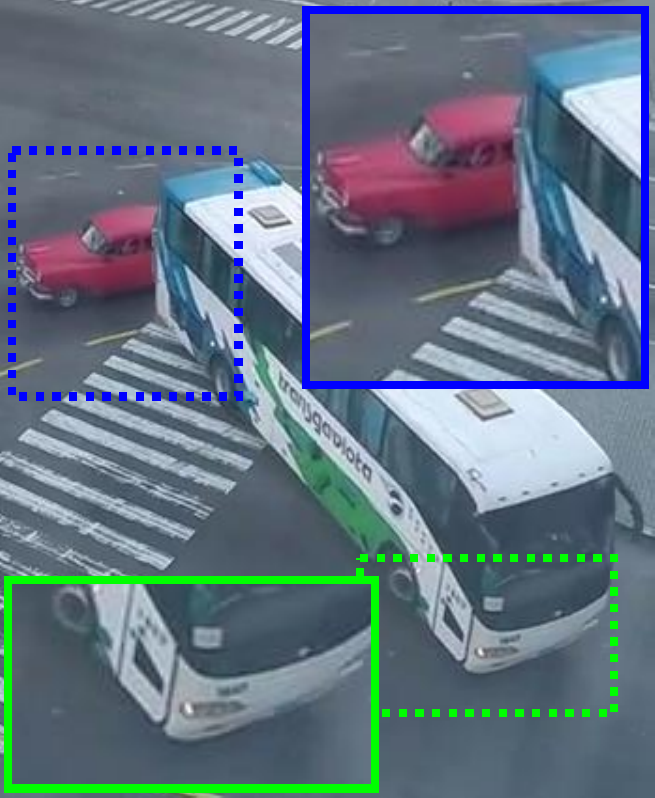}
    	\caption{Ours final}
    	\label{fig:compositedpatch}
	\end{subfigure}
	\begin{subfigure}{0.275\linewidth}
    	\includegraphics[height=1.1in]{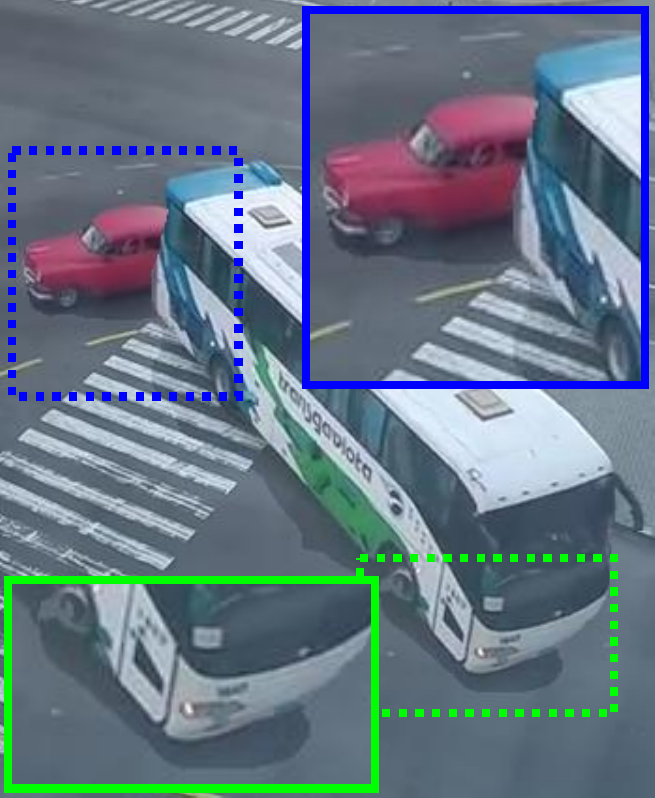}
    	\caption{Thresholded}
    	\label{fig:thresholdedpatch}
	\end{subfigure}
	\begin{subfigure}{0.205\linewidth}
    	\includegraphics[height=1.1in]{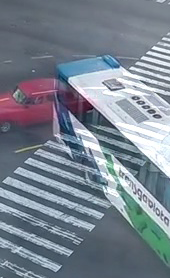}
    	\caption{Mix clone}
    	\label{fig:mixedseamclone}
	\end{subfigure}
	\caption{Result of our user-in-the-loop segmentation procedure and post-process compositing: the raw image patch is placed on the original background (\subref{fig:patchonbg}) and composited using \ch{seamless cloning~\protect\cite{perez2003poisson} to remove lighting changes \wrt BG (red arrows in (\subref{fig:patchonbg})) and our custom algorithm to resolve occlusions (\subref{fig:compositedpatch}). Thresholding the BG difference introduces artifacts (\subref{fig:thresholdedpatch}), while the ``mixed seamless cloning'' in~\protect\cite{perez2003poisson} does not resolve occlusions (\subref{fig:mixedseamclone}). Input~\copyright~Brooks Sherman.}}
	\label{fig:segmentationresult}
\end{figure}

We then use the bounding box to constrain our custom, graphcut-based foreground (FG) segmentation algorithm. Unlike traditional approaches, we aim to composite the patches on their original background (BG). We therefore allow BG pixels to belong to the FG patch as long as \textit{all} FG pixels are correctly classified (see Fig.~\ref{fig:patchonbg}). To ensure this, users can correct any errors in the labeling by interactively scribbling over patches (the colored strokes in (2) in Fig.~\ref{fig:preprocessingui}).

\paragraph{Segmentation algorithm}
After estimating the static background as the per-pixel median of all input frames, we use the seam-finding algorithm in \textit{Graphcut textures}~\cite{kwatra03gctext} to separate FG from BG pixels. We use their pairwise term to conceal seams, and a novel unary term that enforces seam consistency over time and FG pixels to be within the bounding box. Formally, the unary term $U$ for pixel $s$ at position $\mathbf{X}(s)$ belonging to the FG in frame $t$ is defined as
\begin{equation}
\begin{split}
U\left(s, \mathbf{X} \right) = (1-\alpha) &\left[ - \frac{1}{2 \sigma^2} {\Big\|\mathbf{X}(s) - \mathbf{X}_{\mathrm{c}} \Big\|}^{2} \right] \; +  \\
	\alpha &\left[ 1 - M_{t-1}\!\Big( \, \mathcal{F}_{\leftarrow t}\big( \mathbf{X}(s) \big)\Big) \right] \;, 
\end{split}
\end{equation}
where $\mathbf{X}_{\mathrm{c}} = (x_{\mathrm{c}}, y_{\mathrm{c}})$ are the coordinates of the center of the bounding box in image space, $\mathcal{F}_{\leftarrow t}(\cdot)$ is the optical flow function that maps a pixel to its location in the previous frame~\cite{farneback2003two} and $M_{t-1} \in \{0,1\}$ is the pixel mask (FG/BG) of the previous frame $t-1$. We use $\alpha = 0.35$ against a fixed cost to the BG. User-defined scribbles fix pixels' unary cost depending on their association; see Fig.~\ref{fig:segmentationresult} for an example output.

\paragraph{Action definition}
\label{sec:semantics}
The main innovation of our paper is the direct mapping between arbitrary, user-defined, semantic actions and video synthesis commands. Users quickly and intuitively guide our synthesis algorithm towards their goal by issuing these commands; for instance, requesting a candle flame to flicker to the right.
\begin{figure}[ht]
\centering
\includegraphics[width=0.95\linewidth]{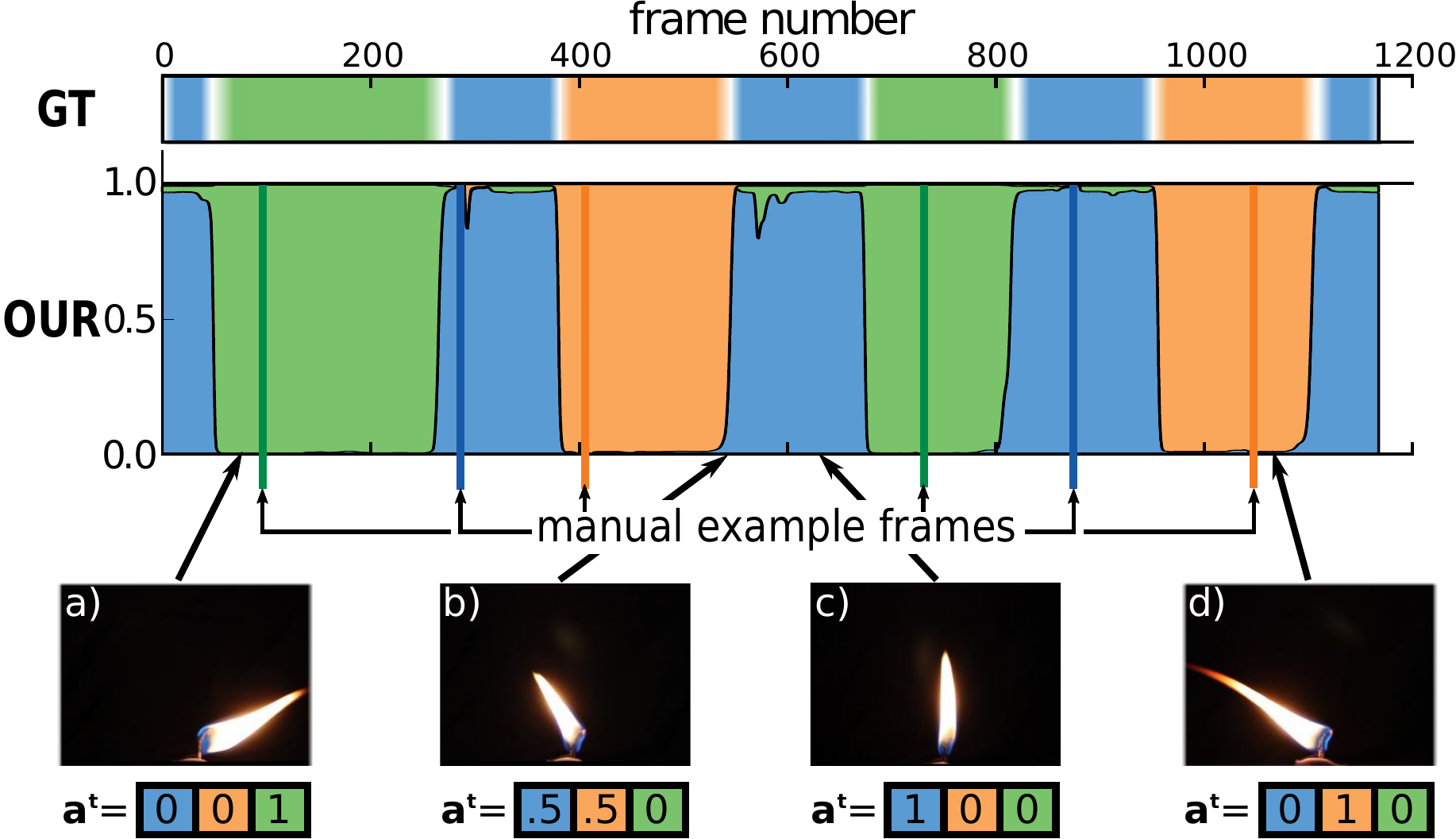}
\caption{
\ch{We show the automatically propagated action assignments (OUR) as opposed to the ground truth (GT). Two examples are given manually (vertical lines) for each one of the three actions (denoted with different colors). The values of the action vector $\mathbf{a}^t$ are shown for 4 example frames. Note how frame b) is correctly ``softly'' assigned to an action between ``\textcolor{orange}{left}'' and ``\textcolor{blue}{rest}'' (not present in GT).}}
\label{fig:labelpropagation}
\end{figure}
In contrast, traditional approaches expect users to manipulate the timelines of several clips by cutting, re-arranging and synchronizing them~\cite{joshi2012cliplets, Lu2012timelineediting}; we believe this makes for far less intuitive and powerful video synthesis.

Action recognition is a well studied problem in the literature. However, existing methods focus on specific use cases, \eg human actions~\cite{weinland2011survey}. Not wanting to impose restrictions, we allow users to indicate actions of interest ``by example'', using our \textit{responsive} interactive tool (Fig.~\ref{fig:preprocessingui}). 
To define actions, users inspect actor sequences (potentially tracked and segmented) and indicate example frames for each action with the press of a button. Users receive \textit{immediate} feedback on the quality of the frame-to-action association of the remaining input frames, as
\chn{they are}
automatically 
\chn{compared}
to the user-given examples.

We can view this as a fuzzy clustering problem, where each action (\eg ``sit'' and ``stand'' for \textsc{Wave} in Fig.~\ref{fig:preprocessingui})
\ch{ is a cluster. In practice, we represent the action visible in frame $t$ of actor sequence $\mathcal{S}$ \chn{for which $l$ distinct actions have been defined} as an $l$-dimensional vector $\mathbf{a}^t$. It represents a probability distribution over the action space, so $||\mathbf{a}^t||=1$. Intuitively, the higher the value of the $l^\mathrm{th}$ element of $\mathbf{a}^t$, the more representative is frame $t$ of the $l^\mathrm{th}$ action.} \chn{For frames indicated as examples of a given action, $\mathbf{a}^t$ takes the form of a binary vector with a $1$ for the specified action and $0$'s elsewhere.} For instance, given the $l=3$ actions defined for \textsc{Candle} (\ie \ch{``\textcolor{blue}{\textbf{rest}}'', ``\textcolor{orange}{\textbf{left}}'' and ``\textcolor{green}{\textbf{right}}'' in Fig.~\ref{fig:labelpropagation}),} a confident example frame showing the flame flickering to the left would be associated the action vector $\mathbf{a}^t = [0, 1, 0]$ \ch{(Fig.~\ref{fig:labelpropagation}d)}.

\chn{We then quickly propagate the user-given information to the remaining frames using~\cite{zhu2003semi}.}
Action vectors \chn{$\mathbf{a}^t$, with $||\mathbf{a}^t||=1$,} are assigned to \chn{all frames, softly clustering them into different actions based on similarity to example frames.} \ch{
The distance between each frame pair $(t, t')$ is defined as}
\begin{equation}
	\label{eq:framedistance}
	D\left(t, t'\right) = \frac{1}{N_\mathrm{O}}\sum_{n=1}^N {\Big[\mathbf{I}\big(t, \mathbf{X}(n)\big) - \mathbf{I}\big(t', \mathbf{X}(n)\big) \Big]}^2 \;,
\end{equation}
where we take the $\mathbf{L}_2$ distance between color intensities $\mathbf{I}\big(t, \mathbf{X}(n)\big)$ and $\mathbf{I}\big(t', \mathbf{X}(n)\big)$ of \ch{every pixel $n$. If the actor sequence has been tracked, we first place the frame's segmented patch onto the static background as shown in Fig.~\ref{fig:patchonbg}. This ensures Eq.~\ref{eq:framedistance} can be used for both tracked and full frame sequences, and spatial relationships are preserved. To avoid bias due to camera-related effects, such as foreshortening, we normalize the distance measure by the number of overlapping pixels $N_\mathrm{O}$ between each frame's bounding box.} We set $N_\mathrm{O}$ to the whole frame area if no bounding box is defined. \ch{For space reasons we do not discuss the propagation further. Please see~\cite{zhu2003semi} and specifically their Eq.(5) for more details.}

Each input frame is associated to an action-cluster ``softly'' as shown in Fig.~\ref{fig:labelpropagation}. This is critical, as frames for which no clear association exists
\chn{(\eg (Fig.~\ref{fig:labelpropagation}b)), }
are used as in-between transitions by our synthesis algorithm. In contrast, traditional video annotation tools, such as ANVIL~\cite{kipp2014anvil}, enable a similar partitioning of video sequences but with hard boundaries between \textit{manually} defined intervals (see Fig.~\ref{fig:labelpropagation} GT), losing the expressiveness of fuzzy assignments in the process.
\begin{figure}[ht]
\centering
\includegraphics[width=\linewidth]{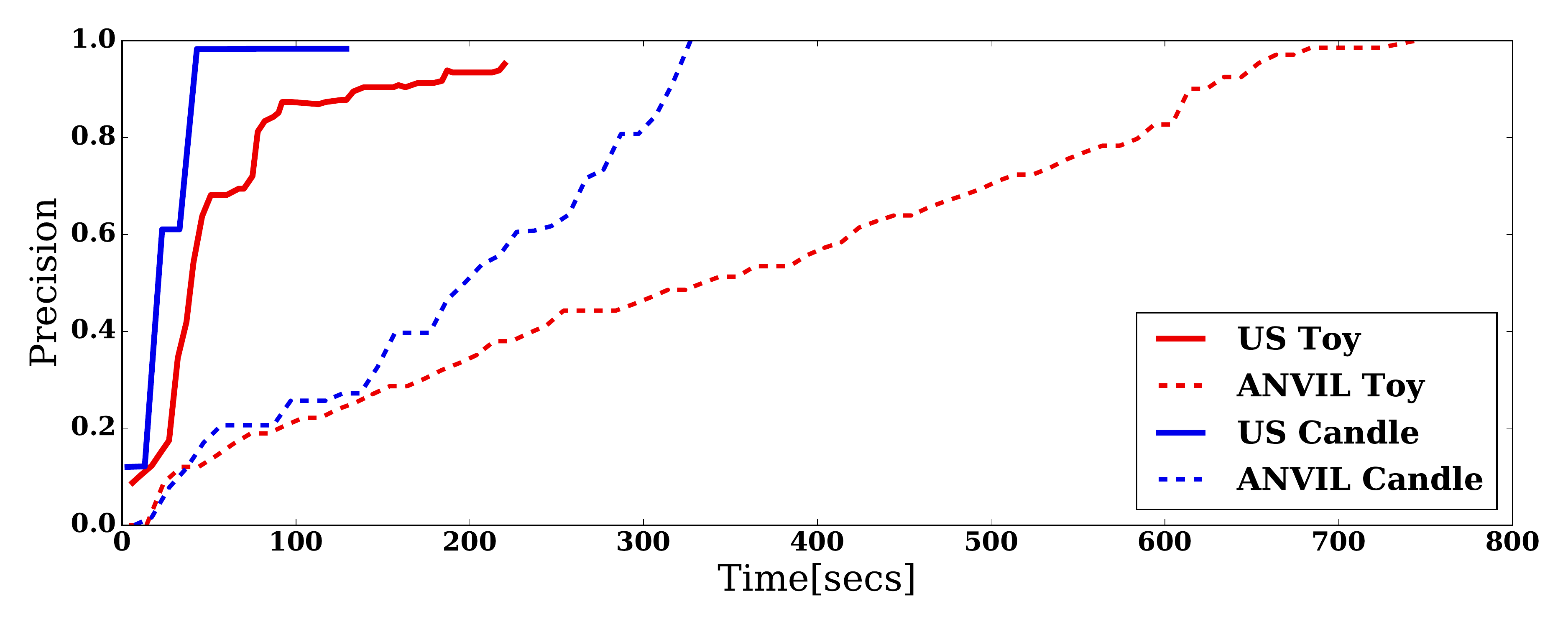}
\caption{Time we spent labeling actions for datasets \textsc{Toy} (700 frames) with 9 different actions and \textsc{Candle} (4000 frames) with 3 different actions using our system or ANVIL~\protect\cite{kipp2014anvil}. Precision computed w.r.t ANVIL labels.}
\label{fig:propagationvsanvil}
\end{figure}
Moreover, as shown in Fig.~\ref{fig:propagationvsanvil}, we experienced a $2\times$ to $3\times$ speed-up in reaching the accuracy permitted by ANVIL (and ignoring in-between frames) 
\chn{thanks to the automatic label propagation from~\cite{zhu2003semi}.}

\section{Video \ch{Performance}}
In this section, \ch{we show how we synthesize a live video performance given} a set of input actor sequences. First, \chn{users} \ch{interactively define the frame compatibility measure, which is later} used to avoid implausible outputs. Second, we present our optimization strategy that \ch{balances} user commands, frame compatibility and transition quality information to synthesize new videos.

\begin{figure}[t]
\centering
	\begin{subfigure}{0.29\linewidth}
    	\includegraphics[width=\linewidth]{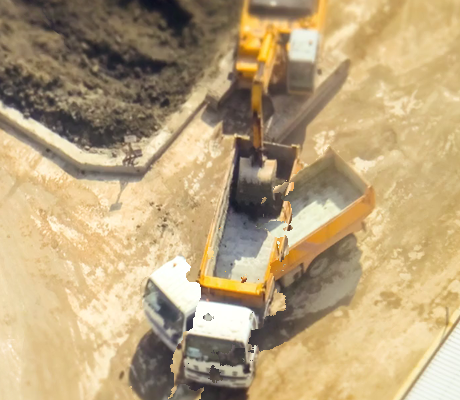}
    	\caption{Without}
    	\label{fig:nocompat}
	\end{subfigure}~~~
	\begin{subfigure}{0.29\linewidth}
    	\includegraphics[width=\linewidth]{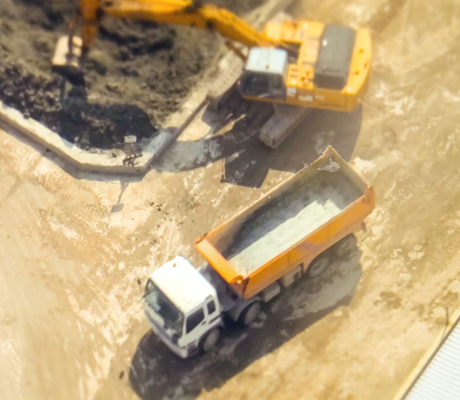}
    	\caption{With}
    	\label{fig:withcompat}
	\end{subfigure}
	\caption{
    Compatibility illustration. Here, the digger is requested to ``load'' a truck while the truck is asked to ``drive'' away in both cases. (\subref{fig:nocompat})~Without frame compatibility, the two actors are free to perform these incompatible actions, with obvious artifacts. (\subref{fig:withcompat})~With \ch{it,} the digger is forced by our algorithm to ``load'' only when the truck actor is ``parked''. \ch{Input~\copyright~Perfect Lazybones/Shutterstock.com}}
	\label{fig:semanticcompatibility}
\end{figure}

\begin{figure*}[t!]
\centering
\includegraphics[width=0.85\linewidth]{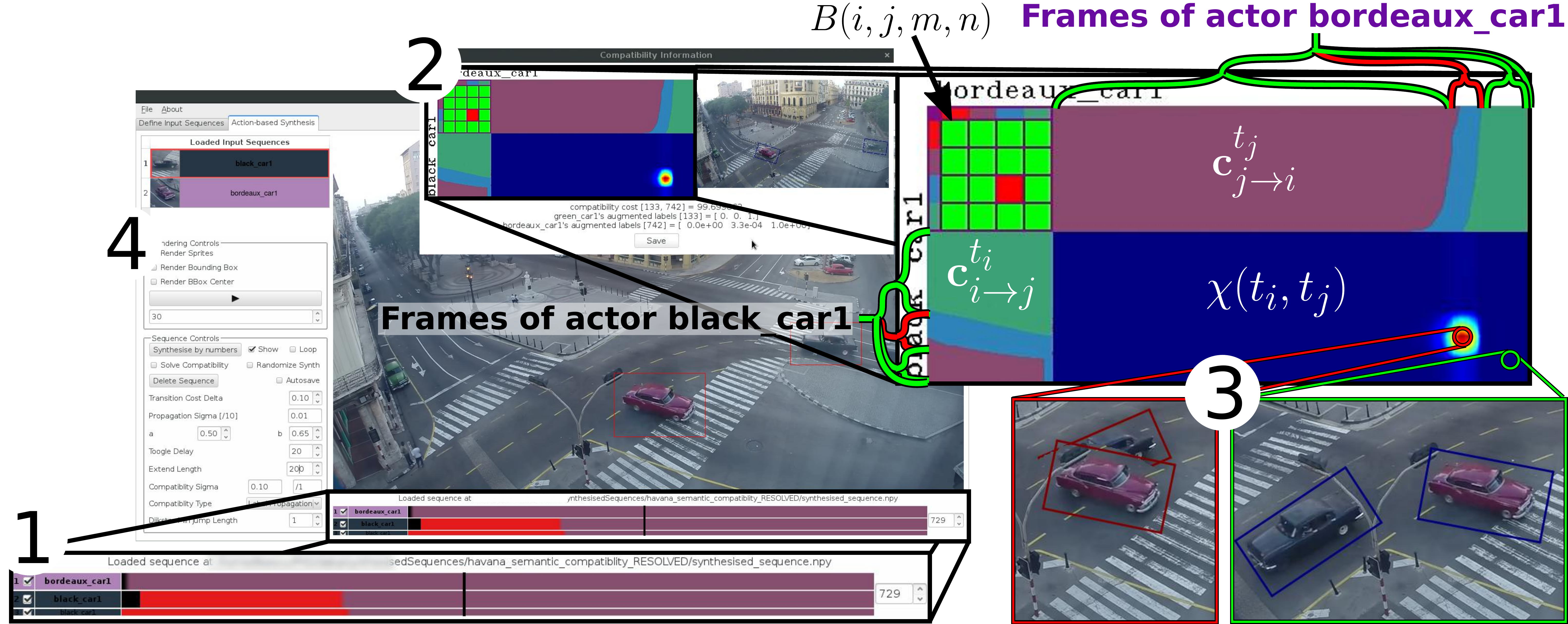}
\caption{Our video synthesis user interface: (1) \textit{output timeline}: lists the used actor sequences (\eg 1 bordeaux car and 2 black cars) and the color coded user-given commands (\eg \textcolor{red}{red} for the car to stay hidden and \textcolor{Fuchsia}{purple} for it to drive through the crossing); (2) \textit{frame compatibility tagging} interface: pairs of frames (previewed to the right of the compatibility info) can be tagged as compatible or incompatible;
(3) \textit{compatibility info} for the two selected actors ($i$ and $j$): frame association to compabtibility clusters per actor ($\mathbf{c}_{i \rightarrow j}^{t_i}$ and $\mathbf{c}_{j \rightarrow i}^{t_j}$), the cluster-pair compatibility ($B(i, j, m, n)$, see Eq.\ref{eq:compatibilityconstant}), for instance, here, the $3^{\mathrm{rd}}$ cluster of each actor (\textcolor{Cerulean}{dark cyan} above) are incompatible as denoted by \textcolor{red}{red} in the cell $(2, 2)$ and the frame compatibility measure $\chi \big( t_i, t_j \big)$ from Eq~\ref{eq:framecompatibilitycost} (\textcolor{blue}{blue} denotes a low cost and \textcolor{red}{red} denotes a high cost) (4) list of available actor sequences (added to the output timeline shown in (1)) and synthesis parameters. \ch{Input~\copyright~Brooks Sherman}}
\label{fig:synthesisui}
\end{figure*}

\paragraph{Frame compatibility}
\label{sec:compatibility}
\ch{As we will see later,} users guide the video synthesis by requesting when actors should perform actions. When multiple actors are present in the same frame \ch{however,} outputs can exhibit implausible situations depending on when users issue their commands \ch{(see \textsc{Havana} cars or \textsc{Candle} flames).}
For instance, \textsc{Candle} flames could flicker in different directions at the same time (Fig.~\ref{fig:candledataset}), a digger could \ch{start loading a moving truck} (Fig.~\ref{fig:nocompat}) or cars could collide in \textsc{Havana} (Fig.~\ref{fig:synthesisui}). In our system, these \textit{incompatibilities} take the form of two actors' frames being composited together onto the output background.

In essence, we again want to assign frames to a set of clusters fuzzily, as we did for our action definition. \chn{These clusters further decompose the actions into sub-sequences. Users mark them as (in)compatible \wrt the sub-sequences of other actors, indicating which sets of frames} 
should be allowed to co-exist in the output video. Given actor sequences $\mathcal{S}_i$ and $\mathcal{S}_j$, we define the compatibility between their respective clusters $m$ and $n$ as
\begin{equation}
	\label{eq:compatibilityconstant}
	B(i, j, m, n) = \left\{ \begin{array}{ll} 1 & \mathrm{if~compatible} \\[5pt] 100 & \mathrm{if~incompatible} \end{array} \right. \; .
\end{equation}
\ch{Initially, \chnn{there is one sub-sequence cluster for each user-defined action, so $m$ and $n$ are in the range $[0, l)$. Later, we discuss how users marking (in)compatibilities changes the number of clusters.}  \chnn{We initialize} 
$B(i, j, m, n)=1$ for all combinations of $m$ and $n$.}
We use $\mathbf{c}_{i \rightarrow j}^{t_i}$ to denote the vector containing the probability that frame $t_i$ of actor $\mathcal{S}_i$ belongs to clusters compatible with actor $\mathcal{S}_j$. Similarly, we define $\mathbf{c}_{j \rightarrow i}^{t_j}$ for frames of actor $\mathcal{S}_j$. We initialize $\mathbf{c}_{i \rightarrow j}^{t_i} = \mathbf{a}^{t_i}$ as it provides an initial division of the input frames, the combination of which could be incompatible.
The compatibility between frame $t_i$ of $\mathcal{S}_i$ and frame $t_j$ of $\mathcal{S}_j$ is then defined as
\begin{equation}
	\label{eq:framecompatibilitycost}
	\chi \big( t_i, t_j \big) = \sum_{m} \sum_{n} \mathbf{c}^{t_i}_{i\rightarrow j}[m] \; \mathbf{c}^{t_j}_{j \rightarrow i}[n] \; B(i, j, m, n) \; ,
\end{equation}
where $\mathbf{c}^{t_i}_{i\rightarrow j}[m]$ denotes the $m^{\mathrm{th}}$ element of $\mathbf{c}^{t_i}_{i\rightarrow j}$. Intuitively, the higher the probability that two frames belong to two compatible clusters, the lower the value of $\chi \big( t_i, t_j \big)$, denoting a low compatibility cost.

Using our GUI (Fig.~\ref{fig:synthesisui}) at synthesis time, users can tag pairs of frames as compatible or incompatible. \ch{Given the pair $\left( t_i, t_j\right)$ we allow two options, which we illustrate for $t_i$ only, as they are analogous for $t_j$:
\begin{enumerate}
	\item \textit{specialize} the compatibility by using $t_i$ as an example for a new cluster $\tilde{m}$,
    \chn{re-running label propagation using the extended set of examples to compute (the now 1D larger) $\mathbf{c}^{t_i}_{i\rightarrow j}$ for all frames of $\mathcal{S}_i$,}
    extending $B$ by one row, setting $B(i, j,\{m \mid m \neq \tilde{m}\}, n) = 1$ and $B(i, j, \tilde{m}, n)$ according to the user input;
	\item \textit{refine} the compatibility measure by leaving $B$ unchanged, setting $t_i$ as a new example for the cluster $m = \underset{m}{\arg\!\max}\big[\mathbf{c}^{t_i}_{i\rightarrow j}\big]$ it most likely belongs to and, 
    \chn{again, re-running label propagation to re-compute $\mathbf{c}^{t_i}_{i\rightarrow j}$.}
\end{enumerate}
If the compatibility of $\left(t_i, t_j\right)$ is changed, we assume option 1, otherwise, the user is asked to decide.}

Intuitively, using the example of the two cars at the crossing in Fig.~\ref{fig:synthesisui}, if one chooses to \textit{specialize}, each cluster will contain frames showing the car in different parts of the intersection. All frames showing the cars in the middle of the crossing can then be marked as incompatible and avoided in the output, making our synthesis continuously smarter.

\paragraph{Action-based video synthesis}
\label{sec:synthesis}
An output video is composed of the frames of one or more actor sequences, re-arranged to \textit{infinitely loop} showing specific \textit{actions} and the \textit{transitions} between them. We phrase our synthesis process as a labeling problem over a two-dimensional graph with $D$ rows and $K$ columns. Each $d$ row is an output layer that contains the frames $\{t_i\}$ of a specific actor sequence $\mathcal{S}_i$, re-arranged to adapt to the user's commands. Each $k$ column represents a final output frame as the union of the frames chosen for each layer. For instance, the \textsc{Toy} result has one row with output frames straight from the input, while the \textsc{Havana} output has as many rows as controllable cars. The label assigned to each $(d, k)$ node is the index of an actor's frame.

Users control the output video by selecting one output layer at a time (see \textit{output timeline} in Fig.~\ref{fig:synthesisui}) and pressing the key associated to the action they want the actor to perform. This in turn defines for each output frame $k$ a $D$-dimensional requested action vector $\{\mathbf{r}^k_d\}$, $d \in [1,D]$. We use a smooth-step function to automatically switch from the currently shown action to the one associated to the user-pressed key stroke.

Formally, our optimization strategy minimizes the energy function
\begin{equation}
	\label{eq:objfunc}
	E = \sum_{k=1}^{K} \sum_{d=1}^{D} \alpha E_{\mathrm{A}} + (1-\alpha) \Big( \beta \, E_{\mathrm{C}} + (1-\beta) E_{\mathrm{T}} \Big) \;,
\end{equation}
where $E_{\mathrm{A}}, E_{\mathrm{C}}$ and $E_{\mathrm{T}}$ are \textit{action}, \textit{compatibility} and \textit{transition costs} respectively. 
Intuitively, the higher the value of $\alpha$, the more responsive the synthesis is to the requested actions (as $E_{\mathrm{A}}$ counts more towards total energy) at the expense of looping quality. This is in turn controlled by the other two terms, balanced by $\beta$. The higher its value, the more important it is to show compatible frames ($E_{\mathrm{C}}$) at the expense of smooth transitions ($E_{\mathrm{T}}$). Both parameters are user-tuned.

We now define the three components of our energy function. For output layer $d$, $E_{\mathrm{A}}$ 
is the cost of showing a frame $t_i^k$ of actor sequence $\mathcal{S}_i$, in output frame k, based on whether the action it shows $\mathbf{a}^{t_i^k}$ matches the requested action $\mathbf{r}^k_d$ and is defined as
\begin{equation}
	\label{eq:actioncost}
	E_{\mathrm{A}}\Big( d, t_i^k \Big) = \frac{1}{2 \sigma_\mathrm{A}^2} {\Big\| \mathbf{a}^{t_i^k} - \mathbf{r}^k_d \Big\|}^2  \;.
\end{equation}
$E_{\mathrm{C}}$ is the cost of showing a pair of frames $t_i^k$ and $t_j^k$, from output layers $d$ and $d'$ respectively, in the same output frame $k$ based on the compatibility cost $\chi(\cdot,\cdot)$ from Eq.~\ref{eq:framecompatibilitycost} and is defined as
\begin{equation}
	\label{eq:compatibilitycost}
	E_{\mathrm{C}}\Big( d, t_i^k \Big) = \sum_{j \in [1, D] \setminus d} \chi\Big(t_{i}^k, t_{j}^k\Big) \; .
\end{equation}
Finally, $E_{\mathrm{T}}$ is the cost of showing actor $\mathcal{S}_i$'s frame $t_i^k$ in output frame $k$ after showing a frame $t_i^{(k-1)}$ in the previous output frame $(k-1)$ and is defined as
\begin{equation}
	\label{eq:appearancecost}
	E_{\mathrm{T}}\Big( d, t_i^k \Big) = \exp\left[\frac{1}{\sigma_\mathrm{T}^2}D\Big( t_i^{(k-1)}, t_i^{k} \Big) \right],
\end{equation}
where $D(\cdot, \cdot)$ is defined in Equation~\ref{eq:framedistance}.

\paragraph{Real-time performance}
The objective in Eq.~\ref{eq:objfunc} can be solved using any discrete energy minimization solver that supports multi-label nodes and arbitrary cost functions, \eg TRW-S~\cite{kolmogorov2006convergent}.
\ch{Unfortunately, TRW-S fails to converge in our experiments and our system \textit{requires} immediate feedback to enable live performances.}

Instead, we propose an iterative approach that minimizes our objective function locally.
We define the new 
energy function 
\begin{equation}
	\label{eq:iterativeobjfunc}
	E'(d) = \sum_{k=1}^{K} E_{\mathrm{U}} + E_{\mathrm{P}},
\end{equation}
and solve it using dynamic programming, \chn{one row $d$ at a time, to synthesize $K$ output frames showing the actor associated to the given row}. The only inter-row energy is 
$E_{\mathrm{C}}$ \chn{(Eq.~\ref{eq:compatibilitycost}), that ensures compatibility between frames of different actors}, which we ``bake'' into the unary term $E_{\mathrm{U}}$. We define it and pairwise term $E_{\mathrm{P}}$ as
\begin{equation}
\begin{split}
	\label{eq:iterativeobjfuncunaryandpairwise}
    E_{\mathrm{U}} = \alpha E_{\mathrm{A}} + (1-\alpha)\beta E_{\mathrm{C}}, \\
    E_{\mathrm{P}} = (1-\alpha)(1-\beta)E_{\mathrm{T}},
\end{split}
\end{equation}
respectively. At each iteration we minimize Eq.~\ref{eq:iterativeobjfunc} for each row in the order the output layers are defined and update $E_{\mathrm{U}}$ to consider the already computed rows. We have found that, one iteration is enough to satisfy our \chn{three} constraints and enables real-time synthesis. \chn{Note that, the result seen during a live performance might differ from the one synthesized using the full set of action requests recorded during it. This is due to the fact that, we can better optimize frame compatibilities and transitions between actions, once we exactly know what each actor will be requested to do and when.} \ch{In our experiments, even long (see Table~\ref{tab:datasetsinfo}) actor sequences such as the people in \textsc{Wave} or the flame in \textsc{Candle} never require more than $500\mathrm{ms}$ to synthesize 400 frames. We further drastically reduce these timings ($10\times$ to $30\times$) by using our compression strategy described below. Typically, we synthesize less than 100 frames in $< 20\mathrm{ms}$ every 2-3 seconds in a live performance scenario, making our system very reactive.}

\paragraph{Optimization Compression}
Jumps are rarely perfect, as pointed out by~\cite{schodl2000vt}, so higher quality outputs involve fewer jumps (\ie the original timeline is followed for as long as possible). With this insight, we speed up our optimization further by synthesizing a subset of the output frames using a subset of the input frames, and ``filling in the blanks'' using subsequent frames. For instance, we can optimize for half the output frames using only every other input frame and gain a $10\times$ speed-up. \ch{We experimented with up to $4\times$ compression, meaning we optimize for every $4^{\mathrm{th}}$ frame, with no visible quality penalty.}

\paragraph{Post-Processing Rendering}
\ch{Our optional post-process first uses seamless cloning by P\'{e}rez~\etal~\cite{perez2003poisson}, to merge each patch to the static background and remove artifacts (red arrows in Fig.~\ref{fig:patchonbg}) due to illumination changes. Then, our custom compositing algorithm resolves occlusions between overlapping segmented actor patches.}

As shown in Fig.~\ref{fig:patchonbg}, our segmented patches often contain background pixels. This was deliberate as it allows us to retain small details such as soft shadows, and works well when patches are placed on their original background. \ch{When patches of different actors overlap in the synthesized frame} (Fig.~\ref{fig:compositedpatch}), BG pixels may obscure FG pixels. \ch{For each pixel where this happens, we dynamically decide which patch is most likely to be FG based on its color intensity difference to the BG.} This approach is more flexible and gives better results than setting a global threshold on the background difference, as shown in Fig.~\ref{fig:thresholdedpatch}. \ch{It is also more suited to our problem than the ``mixed seamless cloning'' in~\cite{perez2003poisson} which does not perform well with complex backgrounds or occlusions, and introduces ghosting (Fig.~\ref{fig:mixedseamclone}).}

\section{Creative synthesis}
In contrast to existing tools, our system accepts high-level, user-defined commands that guide our video synthesis algorithm. Users need simply request when they want an actor to perform an action, enabling many alternative and fun ways of creating videos, \chn{which we now present.}

\begin{figure}[t]
\centering
	\begin{subfigure}{0.37\linewidth}
    	\includegraphics[width=\linewidth]{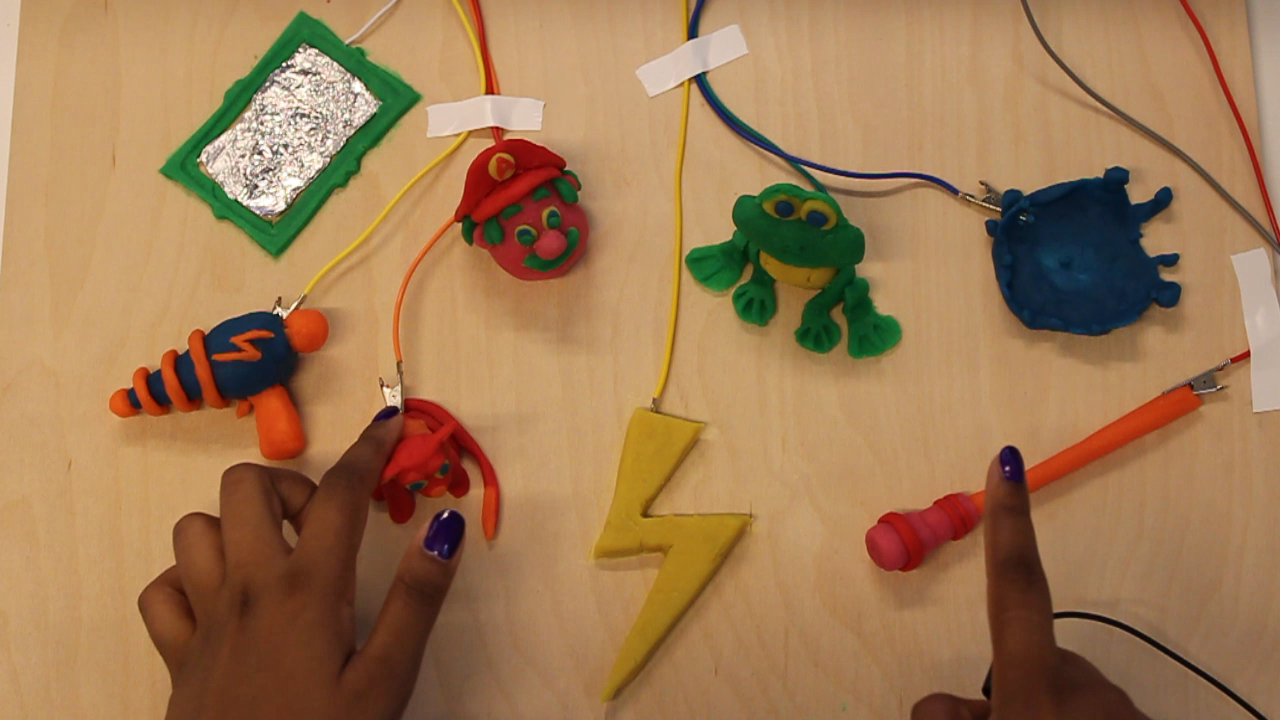}
    	\caption{MakeyMakey~\cite{makeymakey}}
    	\label{fig:makeymakey}
	\end{subfigure}
	\begin{subfigure}{0.37\linewidth}
    	\includegraphics[width=\linewidth]{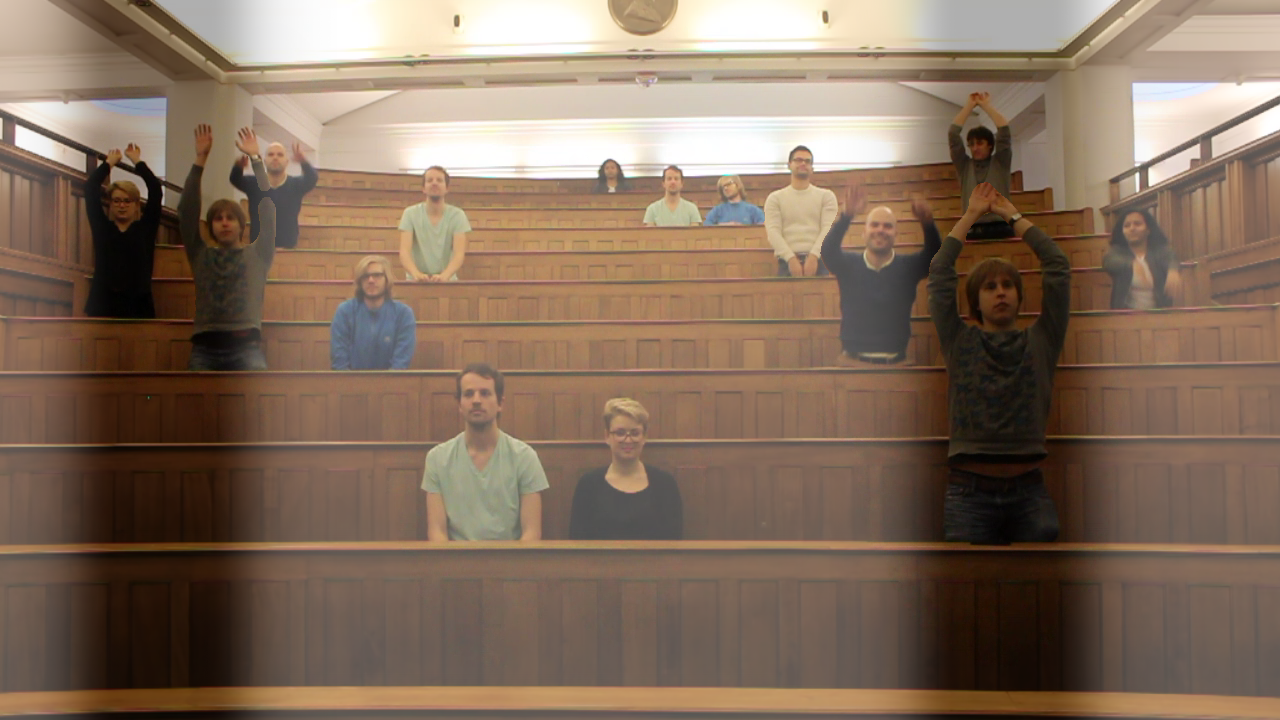}
    	\caption{By numbers}
    	\label{fig:synthbynums}
	\end{subfigure}
	\begin{subfigure}{0.24\linewidth}
    	\includegraphics[width=\linewidth]{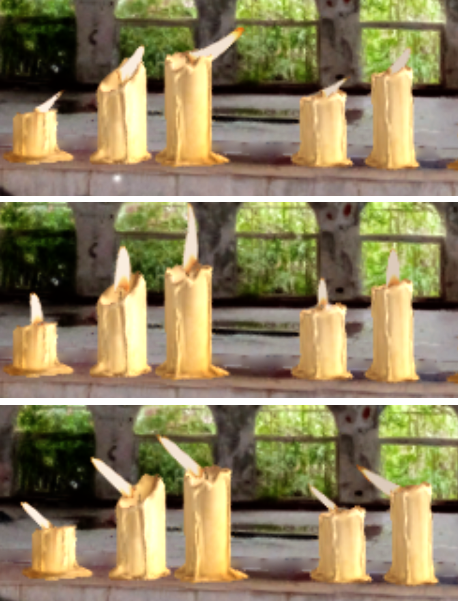}
    	\caption{Game logic}
    	\label{fig:gamelogic}
	\end{subfigure}
	\caption{Using actions as an abstraction for synthesis commands enables new and creative ways of creating videos. Trigger commands can be given through touching a keyboard or \textit{Play-doh} figurines (\subref{fig:makeymakey}), animated color bars (\subref{fig:synthbynums}) or context specific game logic (\subref{fig:gamelogic}).}
	\label{fig:creativesynthesis}
\end{figure}

\paragraph{Keyboard and MakeyMakey}
The simplest way to create a video with our system is by using a keyboard. An action for each actor can be mapped to a specific key stroke which, when pressed, signals our synthesis algorithm to show frames from that category. Given the simplicity of our mapping process, we can use more creative input methods too. For instance, in Fig.~\ref{fig:makeymakey} and in our supplemental video, an artist uses MakeyMakey~\cite{makeymakey} and some \textit{Play-doh} figurines to create a video using our \textsc{Drumming} dataset, where specific drums or cymbals are hit when the associated figurine is touched.

\paragraph{Synthesis by numbers}
Our system enables creation of video analogous to image synthesis~\cite{hertzmann2001image}. We associate actions to solid colors, and create an animated control sequence showing those colors using any paint tool. Actions are then triggered according to the colors shown by the control sequence. For instance, Fig.~\ref{fig:synthbynums}, shows an animated black bar crossing the screen from left to right. At each output frame, people in \textsc{Wave} are asked to ``stand'' if they are under the black bar and ``sit'' otherwise. This allows us to quickly create a Mexican Wave. In our supplemental video we show that we can easily change the control sequence to quickly 
\chn{synthesize completely different waves.}

\paragraph{Game Logic}
Our system also allows external factors to drive video synthesis. In particular, custom video-game logic can be programmed to issue commands to our synthesis algorithm based on dynamic game-related events. For instance, we have 
\chn{embedded a pre-computed set of outputs of our controllable \textsc{Candle} into a game level (see Fig.~\ref{fig:gamelogic} and supplemental video).}
Then, the game logic decides how the candle should react to its own wind simulation, by for instance, making it flicker to the left or to the right.

\section{Results}
\label{sec:results}
\begin{figure*}[t!]
	\centering
	\begin{subfigure}{0.12\linewidth}
    	\begin{subfigure}{\linewidth}
        	\includegraphics[width=\linewidth]{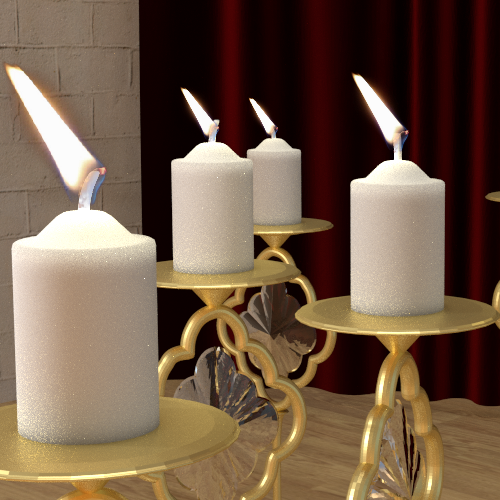}
    	\end{subfigure}
    	\caption{\textsc{Candle}}
    	\label{fig:candledataset}
	\end{subfigure}
	\begin{subfigure}{0.12\linewidth}
    	\begin{subfigure}{\linewidth}
        	\includegraphics[width=\linewidth]{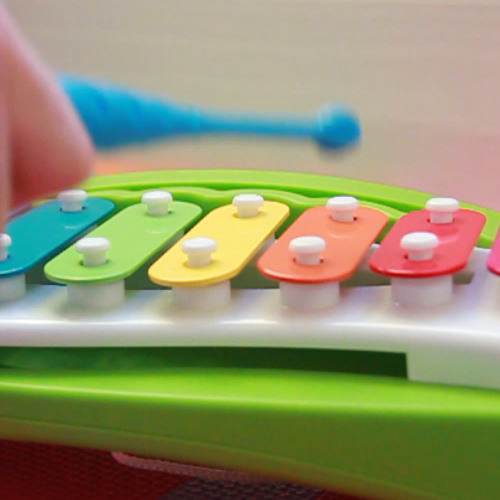}
    	\end{subfigure}
    	\caption{\textsc{Toy}}
    	\label{fig:toydataset}
	\end{subfigure}
	\begin{subfigure}{0.12\linewidth}
    	\begin{subfigure}{\linewidth}
        	\includegraphics[width=\linewidth]{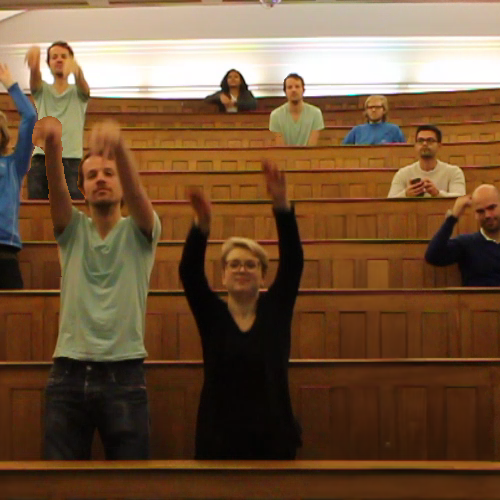}
    	\end{subfigure}
    	\caption{\textsc{Wave}}
    	\label{fig:wavedataset}
	\end{subfigure}
	\begin{subfigure}{0.12\linewidth}
    	\begin{subfigure}{\linewidth}
        	\includegraphics[width=\linewidth]{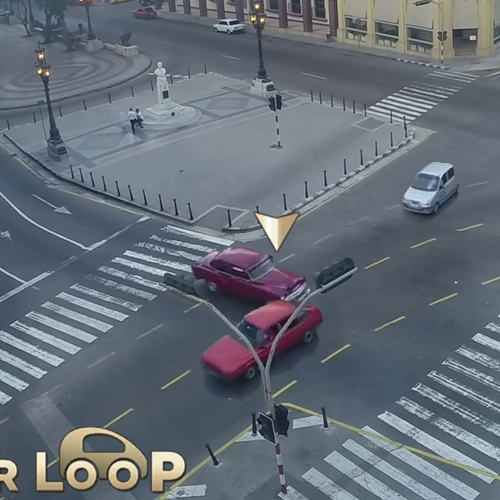}
    	\end{subfigure}
    	\caption{\textsc{Havana}}
    	\label{fig:havanadataset}
	\end{subfigure}
	\begin{subfigure}{0.12\linewidth}
    	\begin{subfigure}{\linewidth}
        	\includegraphics[width=\linewidth]{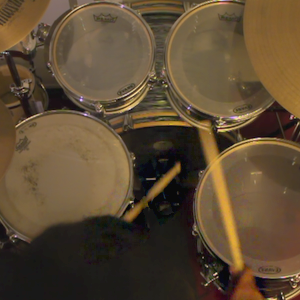}
    	\end{subfigure}
    	\caption{\textsc{\ch{Drumming}}}
    	\label{fig:drummingdataset}
	\end{subfigure}
	\begin{subfigure}{0.12\linewidth}
    	\begin{subfigure}{\linewidth}
        	\includegraphics[width=\linewidth]{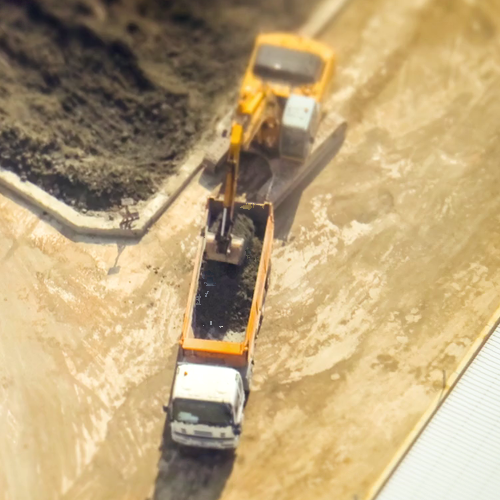}
    	\end{subfigure}
    	\caption{\textsc{Digger}}
    	\label{fig:diggerdataset}
	\end{subfigure}
	\begin{subfigure}{0.12\linewidth}
    	\begin{subfigure}{\linewidth}
        	\includegraphics[width=\linewidth]{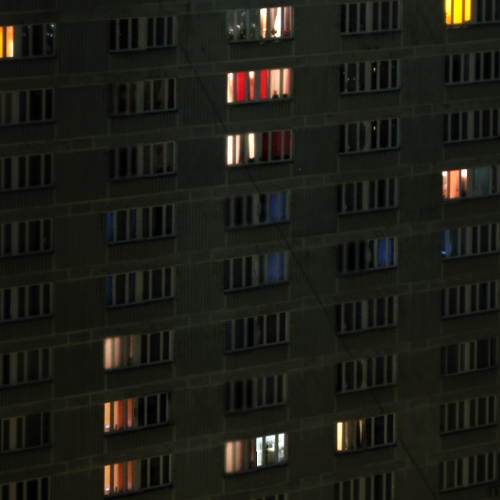}
    	\end{subfigure}
    	\caption{\textsc{Windows}}
    	\label{fig:windowsdataset}
	\end{subfigure}
	\begin{subfigure}{0.12\linewidth}
    	\begin{subfigure}{\linewidth}
        	\includegraphics[width=\linewidth]{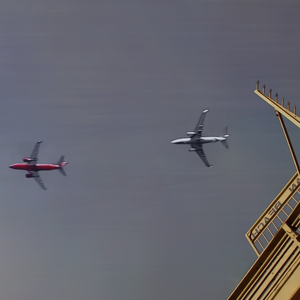}
    	\end{subfigure}
    	\caption{\textsc{Planes}}
    	\label{fig:planesdataset}
	\end{subfigure}
	\caption{
    Sample output frames.
    \ch{Inputs to (\subref{fig:havanadataset})~\copyright~Brooks Sherman, 	(\subref{fig:diggerdataset})~\copyright~Perfect Lazybones/Shutterstock.com, (\subref{fig:windowsdataset})~\copyright~Pavel L/Shutterstock.com, (\subref{fig:planesdataset})~\copyright~Cysfilm. Results sequences shown in supplemental video.}
    }
	\label{fig:datasets}
\end{figure*}

The new medium of expression described in this paper enables the creation of a wide variety of video performances. To stimulate the reader's creativity, we and our users have produced a number of output videos using the system. They can be seen in our supplemental video\footnote{\label{fn:website}Visit our website \href{http://visual.cs.ucl.ac.uk/pubs/actionVideo/}{http://visual.cs.ucl.ac.uk/pubs/actionVideo/}}, as stills in Fig.~\ref{fig:datasets}, and are briefly described here. In Table~\ref{tab:datasetsinfo}, we provide information about the \chn{actors defined for each dataset and the needed user effort.}

We use \textsc{Candle} as a didactic example. After segmenting the flame using pixel intensity, we define three actions (``left'', ``right'', ``rest'') and thus are able to have it react according to a hypothetical breeze. Given multiple copies of a candle, as shown in our supplemental video, we also tag pairs of frames showing distinct actions as incompatible, such that our synthesis algorithm can ensure they all react to the breeze randomly, but in the same manner, without having to manually ensure it for each flame. Similar results 
\ch{are achieved} from within a videogame level, as shown in Fig.~\ref{fig:gamelogic}.

\paragraph{Range}
\begin{table}
  \centering 
  \begin{tabular}{l r r r r r}
    {\small\textit{Dataset}}
    & {\small \textit{Actors}}
      & {\small \textit{Actions}}
     & \begin{tabular}{@{}c@{}}\small \textit{Average} \\ \small \textit{\#frames}\end{tabular}
     & \begin{tabular}{@{}c@{}}\small \textit{Avg Prep} \\ \small \textit{[s/actor]}\end{tabular}
     & \begin{tabular}{@{}c@{}}\small \textit{Output} \\ \small \textit{layers}\end{tabular}\\
    \midrule
    \textsc{Candle} & 1 & \{3\} & 1168 & 60 & 8 \\
    \textsc{Toy} & 1 & \{9\} & 702 & 210 & 1 \\
    \textsc{Wave} & 18 & \{2\} & 1124 & 1320 & 15 \\
    \textsc{Havana} & 13 & \{2\} & 587 & 1257 &  \\
    \textsc{Theme Park} & 13 & \{1, 2\} & 630 & 820 & 21 \\
    \textsc{Digger} & 2 & \{2\} & 160 & 405 & 2 \\
    \textsc{Windows} & 23 & \{2\} & 115 & 60 & 54 \\
    \textsc{Planes} & 7 & \{2\} & 105 & 917 & 10 \\
    \textsc{Drumming} & 3 & \{2, 4, 5\} & 506 & 2440 & 3 \\
  \end{tabular}
  \caption{Example input videos. Note how some datasets contain actors with different numbers of actions. For each dataset, \ch{we show the average number of frames per actor, average number of seconds necessary to prepare them and} the number of layers in the \ch{corresponding} output video.}
    ~\label{tab:datasetsinfo}
\end{table}
\textsc{Havana} and \textsc{Theme Park} show the flexibility of our method and its ability to avoid incompatibilities. These and the subsequent examples differ from \textsc{Candle} because they cannot or would be very hard to make using existing tools. Multiple moving elements were tracked and segmented, and associated to the actions ``visible'' and ``invisible''. Thanks to our frame compatibility measure, we are able to avoid collisions between cars and people when they are visible at the same time in the output video. Similarly, in \textsc{Digger} the user ensured a digger only loads a truck when it is parked, by tagging a few incompatible pairs of actor-frames where the truck is moving while the digger tries to load it.

With the remaining datasets, we showcase further creative interactions with our system. Using \textsc{Wave}, we create a Mexican Wave simply by creating a control animation as seen in Fig.~\ref{fig:synthbynums}. We can then quickly alter the result by simply changing the control sequence, to add a second subsequent wave, one in the opposite direction, or even an interlaced one. In \textsc{Drumming}, we can control a drummer playing his instrument by simply touching \textit{play doh} figurines representing funny sounds, while with \textsc{Toy} we can create a video showing specific songs being played onto a colorful xylophone after filming random notes being hit. \textsc{Windows} allows us to map windows on a building facade to pixels in a grid, and render a compelling game of Tetris by manipulating the light switches. With \textsc{Planes}, airplanes take off in sync with a user hitting the spacebar to the rhythm of \ch{a well known videogame theme-song.} Finally, a game-developer used \textsc{Havana} to create the CounterLoop videogame \chn{(see project webpage).}
All these outputs and use-cases are prepared with the same workflow, qualitatively demonstrating its range.

\section{Empowerment Evaluation}
\label{sec:evaluation}
As they are our closest competitors, we aim to replicate our Mexican Wave output (Fig.~\ref{fig:wavedataset}) using either of~\cite{liao2013,Lu2012timelineediting,joshi2012cliplets}. Liao~\etal~\cite{liao2013} can deal with complicated scenes but it merely finds the best possible looping patches. It does not allow the user to choose where to loop or when people should sit and when to stand. The system of Lu~\etal~\cite{Lu2012timelineediting} can allow us to splice together different sub-clips and re-arrange them. However, they create unnatural speed-ups or slow-downs when people need to sit for longer or less than the input video, because of their time scaling algorithm, and do not account for transitions between the clips as our looping does automatically. 

\ch{We informally compare our system to \textit{Cliplets}~\cite{joshi2012cliplets} by recreating an 8-actor Mexican Wave (see supplemental video) and one candle flame using \textsc{Candle}. Similar to~\cite{Lu2012timelineediting}, \textit{Cliplets} works by defining, manipulating, and arranging layers of video clips to create the output.
\chn{Each layer shows one looping animation (\eg an actor performing one action) or input frames as captured (which we used to transition between actions of the same actor). For instance, a video of a flame flickering left and then right, requires three layers: a) ``loop flame left'', b) ``playback flame going from left to right'', c) ``loop flame right''. Users must manually define when to show each loop and its length and find transition frames in b) such that there is no visible jump when hiding layer a) to show b) and when hiding b) to show c). This very time consuming process is slowed further if the result needs changes, as changing looping time or animation order requires carefully re-arranging layers and redefining transitions, effectively starting over.}
In contrast, our system enables live performances after a one-off preparation stage. The output is created as an endless stream and the user is free to play-act and improvise in real time, \chnn{an invaluable ability unique to our system}. 
\chn{Using \textit{Cliplets}, it took us $4\times$ and $9\times$ longer to recreate \textsc{Wave} and \textsc{Candle} respectively.} \chnn{This is mainly due to manually inspecting the video to find the right subset of frames to loop through or use as transitions between animations.}
}

For reasons discussed above, several methods~\cite{joshi2012cliplets,liao2013,Lu2012timelineediting} were not able to successfully recreate our outputs. Separately from range, we want to assess whether our action-based synthesis \emph{empowers} users' creativity~\cite{DanOlsenCHITalk} and helps express it better and faster than baselines. We therefore compare against Adobe After Effects (AE) because, with proper training, \ch{it} gives users commercially-accepted tools that should reproduce our results. We gathered 6 novice users, \ch{that had never used either system,} and asked them to recreate the Mexican Wave sequence. In particular, they were instructed to create a left-to-right wave, some idle animation in the middle (\ie sitting people) and a final right-to-left wave. 
After relevant training, half of the users used our system while the remaining half performed the same task using AE. \ch{Both sets of users were given the same 7 sprites as input that had already been tracked and segmented.}

\begin{figure}[t]
\centering
\includegraphics[width=\linewidth]{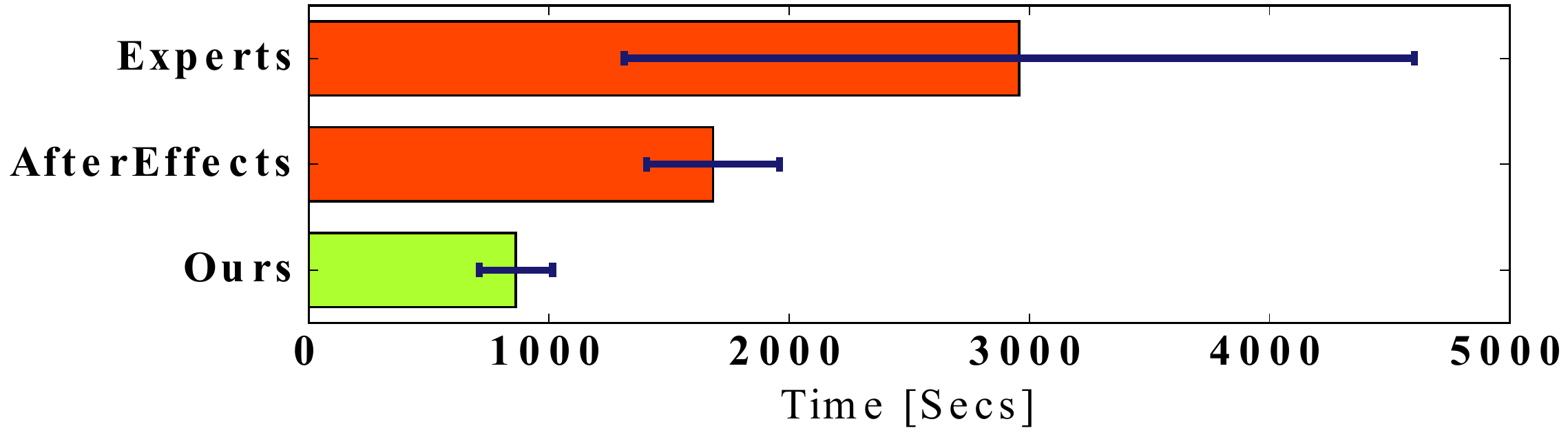}
\caption{Average timings necessary to replicate a variation of our Mexican Wave result. From top to bottom: \textit{expert} users of NukeStudio, Blender, and AfterEffects; novice users of \textit{AfterEffects}; novice users of \textit{our} system.}
\label{fig:userstudy}
\end{figure}

\chn{Fig.~\ref{fig:userstudy} shows that} users of our system were roughly twice as fast, indicating that our system is indeed easy to use. In their words, they really enjoyed the simplicity with which actions are defined, the responsive visualization (Fig.~\ref{fig:preprocessingui}), and the immediate video feedback that comes with action requests during synthesis. \chn{In the supplemental video, we qualitatively show that the AE results are inferior to ours.} This is because our system automatically finds the best transitions between actions, while the AE users need to manually align the clips showing the sitting and standing actions and decide when to transition between them. Finally, we also asked three expert Nuke Studio, Blender, and AE users to recreate the same sequence. Their timings (also in Fig.~\ref{fig:userstudy}) show a larger variability depending on their willingness to find optimal-looking jumps. In fact, they were given the same inputs and task as the novice AE users, but no further instructions, and their results are of varying quality.

\section{Discussions with artists}
\begin{figure}[t]
\centering
\includegraphics[width=0.5\linewidth]{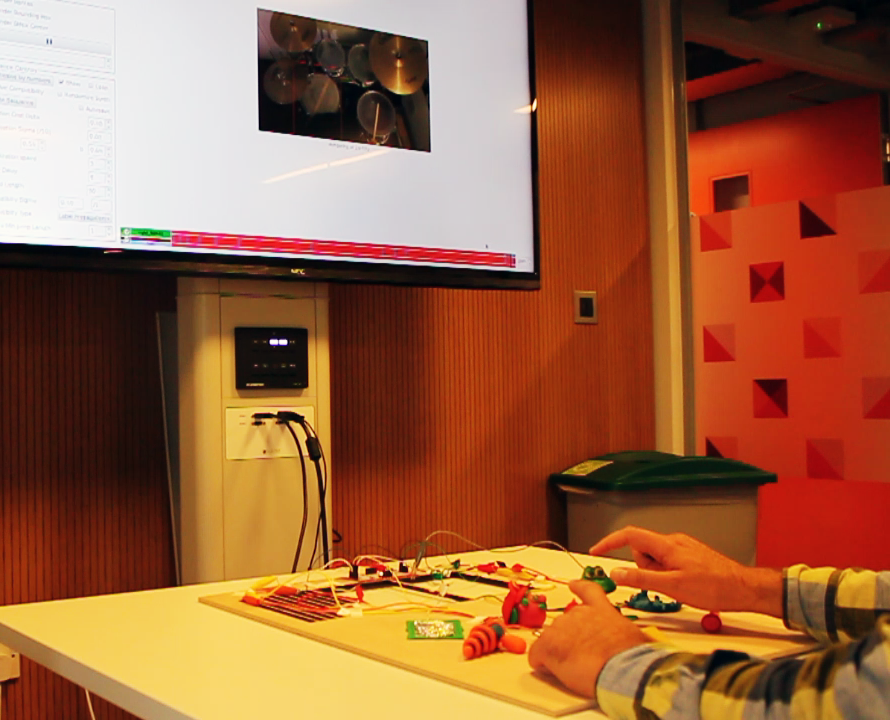}
\caption{\ch{An artist using our live performance system.}}
\label{fig:artist}
\end{figure}

To assess the \textit{balanced} structure~\cite{DanOlsenCHITalk} of our system, we informally interviewed three digital artists, mostly involved with game development or live performance design, and introduced them to our system (see \ch{Fig.~\ref{fig:artist}} using \textsc{Drumming}). They agreed that our system takes a large step toward making ``video composition more like playing a musical instrument'', enabling live performances with immediate video feedback, such as seen in \textsc{Drumming}. We were surprised that one asked to sacrifice video quality for better responsiveness, especially if sound feedback is present. \ch{As shown in our supplemental video,} we were able to cater to this request \ch{by favoring $E_\mathrm{A}$(Eq.~\ref{eq:objfunc}) at the expense of good looking transitions. The result} can then be improved, immediately after recording the user commands, by re-synthesizing the sequence with the default video settings.

Interestingly, \ch{artists} saw our system as a ``sketching tool'' for quick prototyping, such as seen using \textit{synthesis by numbers} to create the variations of the Mexican \textsc{Wave}. In fact, experimenting with choreographies was a suggested use case, such as filming dancers improvising and re-arranging their moves using our system after the fact. They also expressed the desire to have the synthesis algorithm as part of game engines, as they feel it gives them important control over sprite synthesis. When shown our \textsc{Candle} video, they immediately recognized its value for game development and suggested further content, such as water drops into puddles. Finally, they suggested a number of ``shared experiences'' for teaching and training that our system would make possible. For example, a trainer could decide which exercises people should perform and give them live video tutorials, or could trigger traffic scenarios.

\section{Conclusion}
\label{sec:conclusion}
We presented a system that facilitates a new medium of expression, where videos are created much like live looping is composed. Users define actors, optionally tracking and segmenting them, and associate actions to triggers, which can take the form of multiple interfaces. Our workflow helps both novices and advanced users to prepare their footage and, \ch{for the first time}, turn it into interactive \ch{live} performances. 

\paragraph{Limitations}
The quality of our results, ultimately depends on the input videos and the users' willingness to invest the necessary effort to process them. The longer the actor sequences, the greater the variability and coverage of situations. For instance, there are no clean transitions between hitting some notes in \textsc{Toy} and the rest position, because the mallet hand rarely leaves the view in the input video, resulting in \ch{occasional} jumpy animation. \ch{In general,} this holds for short videos where there is too much variability but not enough coverage. This problem could be tackled with interpolation and morphing techniques similar in spirit to~\cite{sevillalara2015smoothloops}. Additionally, there is always a trade-off between how quickly the synthesis shows the desired action, and how smooth the transition looks. Being able to successfully camouflage bad jumps would reduce the time necessary to transition between actions. It would also remove the input lag between the button press and the on-screen response, critical for live performances such as \textsc{Drumming}.

\section{Acknowledgments}
\ch{We would like to thank Mike Terry for his invaluable feedback. Thanks to all the volunteers and artists for their availability during our validation phase, especially Tobias Noller, Evan Raskob and Ralph Wiedemeier. The authors are grateful for the support of EU project CR-PLAY (no 611089) \href{www.cr-play.eu}{www.cr-play.eu} and EPSRC grants EP/K023578/1, EP/K015664/1 and EP/M023281/1.}

\bibliographystyle{SIGCHI-Reference-Format}
\bibliography{sample}

\end{document}